\documentclass[onecolumn,12pt,lettersize]{IEEEtran}
\usepackage[top=1 in, bottom=1 in, left=0.75 in, right=0.75 in]{geometry}

\usepackage[cmex10]{amsmath} 
\usepackage{amssymb}
\usepackage{tabularx}
\usepackage[sc,osf,noBBpl]{mathpazo} 
\usepackage{cite} 
\usepackage[pdftex]{graphicx} 
\usepackage{subfigure}
\usepackage{algorithm}
\usepackage[noend]{algorithmic}
\usepackage{amssymb} 
\usepackage{graphicx}
\usepackage{epstopdf}

\usepackage{stfloats}
\usepackage{tikz}

\usepackage{amsthm} 
\newtheorem{pb}{Problem}
\newtheorem{theorem}{Theorem}
\newtheorem{corollary}{Corollary}
\newtheorem{lemma}{Lemma}

\newtheorem{definition}{Definition}

\newtheorem{assumption}{Assumption}

\begin{document}
\title{On Low-Complexity Quickest Intervention of Mutated Diffusion Processes Through Local Approximation}
\author{Qining Zhang\textsuperscript{1}\thanks{1:Department of EECS, University of Michigan, Ann Arbor, MI 48105, USA}, 
Honghao Wei\textsuperscript{1}, 
Weina Wang\textsuperscript{2}\thanks{2:Computer Science Department, Carnegie Mellon University, Pittsburgh, PA 15213, USA},
Lei Ying\textsuperscript{1}}
\maketitle

\begin{abstract}
  We consider the problem of controlling a mutated diffusion process with an unknown mutation time. The problem is formulated as the quickest intervention problem with the mutation modeled by a change-point, which is a generalization of the quickest change-point detection (QCD). Our goal is to intervene in the mutated process as soon as possible while maintaining a low intervention cost with optimally chosen intervention actions. This model and the proposed algorithms can be applied to pandemic prevention (such as Covid-19) or misinformation containment. We formulate the problem as a partially observed Markov decision process (POMDP) and convert it to an MDP through the belief state of the change-point. We first propose a grid approximation approach to calculate the optimal intervention policy, whose computational complexity could be very high when the number of grids is large. In order to reduce the computational complexity, we further propose a low-complexity threshold-based policy through the analysis of the first-order approximation of the value functions in the \emph{``local intervention''} regime. Simulation results show the low-complexity algorithm has a similar performance as the grid approximation and both perform much better than the QCD-based algorithms. 
\end{abstract}

\section{Introduction}
Mutated diffusion processes such as variants of Covid-19, malicious behaviors in communication networks, and manipulated information on online social networks can create serious public health issues and severe financial crises, with catastrophic social and economic consequences. These highly consequential information and virus ``mutations'' often occur in subtle and sometimes random ways, which makes it difficult to control/intervene at their early stages. For example, the Delta variant of Covid-19, first detected in March 2021, has quickly become the dominant strain in many countries; malicious devices may merely deviate subtly from allocated protocols at the early stage; and misinformation is often embedded in real news by manipulating only a few but not all details of a true news story. The focus of this paper is to develop both theories and algorithms to understand {\em when and how to intervene in a diffusion process, which may have mutated, to control its damage?} 

In this paper, we consider a model where an agent receives a sequence of observations assumed to be drawn from a distribution. Here, an observation may represent the symptom of a Covid-19 patient or whether an online platform user has retweeted the news after reading it on social media. A mutation that occurs at a time unknown to the agent changes the observation distribution, while the sample space remains the same. At each time $t$, given the sequence of observations, the agent decides whether to intervene to control the mutated diffusion process. If so, how strict the intervention action should be to limit its impact, e.g., whether to suggest or mandate mask-wearing, and whether to ask voluntary quarantine or impose a strict lockdown.

We call this problem ``quickest intervention'' because it is closely related to quickest change detection (QCD), where the goal is to quickly detect a change-point based on sequential observations. The QCD problem was first proposed and studied by Shiryev~\cite{shiryaev1963optimum}, where he studied a parameterized observation distribution with a geometric prior for the change-point and proposed the well-known Shiryev's test. Lorden extended Shiryev's theory with the minimax criterion by replacing the geometric prior with any unknown non-random change-point~\cite{lorden1971procedures}. He also showed the CuSum test from~\cite{page54} is asymptotically optimal as false alarm requirements become stricter, while the exact optimality is proved by Moustakides~\cite{moustakides1986optimal} and Ritov~\cite{ritov1990decision}. Improvements based on CuSum have been proposed in various works including~\cite{pollak1985optimal}. Our work is closely related to the Bayesian regime summarized in Shiryev's book~\cite{shiryaev2007optimal}. However,  QCD focuses only on detecting the change-point and does not consider taking actions that will change the distribution of the observations. A recent paper~\cite{Banerjee} does consider intervention actions in QCD, but the problem is studied mostly via simulations. In terms of applications, \cite{WeiKanWan_19} studied the quickest detection of misinformation using the optimal stopping theory, which again does not consider the intervention. 

Since an action changes the underlying diffusion process in ``quickest intervention'', there are fundamental differences between our problem and pure detection. The problem in fact falls into the category of partially observed Markov decision process (POMDP) \cite{krishnamurthy2016partially}, where an agent makes decisions while receiving sequential observations and inferring unobserved states (change-point or mutation). It is well-known that solving general POMDPs is extremely difficult. Only limited structural results and low-complexity solutions with provable performance for general POMDPs are studied. For example, \cite{LovjoyMonotone} established the monotonicity of value functions under the monotone likelihood ratio (MLR) ordering condition. \cite{KrishnamurthyPolicy} later improved the result by developing a low-complexity myopic policy. However, in general, even establishing sufficient conditions that the optimal policy of a POMDP possesses a threshold structure is extremely hard. In this paper, we will first formulate the quickest intervention problem as a POMDP and then develop low-complexity algorithms and theoretical results in the ``local intervention'' regime.

The main contributions of this paper are summarized below:

\textbf{(1) Problem Formulation:} We formulate the quickest intervention problem as a discrete-time process with an unknown change-point (mutation) and multiple intervention actions that can be used to control the diffusion process. The objective of the problem is to identify a policy that minimizes the total propagation and intervention costs. This formulation is analogous to QCD but involves decision-making that affects the underlying diffusion process. 

\textbf{(2) POMDP Approach:} The problem is actually a POMDP where the hidden state is the change-point. We convert it into a fully observed MDP via belief states. Grid approximation is then proposed to compute the optimal policy numerically.

\textbf{(3) \emph{``Local Intervention''} Regime:} To overcome the computational complexity of grid approximation when the number of grids is large, the key contribution of this paper is to develop a low-complexity algorithm based on the first-order approximation of value functions in the ``local intervention'' regime. With a derived approximated Bellman equation, we prove that the action-value function in the ``local intervention'' regime is nearly submodular, and the policy that solves the approximated Bellman equation has a threshold structure. Furthermore, we derive upper bounds on the thresholds with closed-form expressions. Based on the upper bounds, we propose a low-complexity algorithm that increases the intervention levels when the belief exceeds the corresponding threshold. Interestingly, our numerical results show that the low-complexity algorithm is close to optimal even beyond the \emph{``local intervention''} regime, which demonstrates its effectiveness (nearly-optimal performance) and efficiency (very low computational complexity).

\section{Motivating Examples}\label{sec:Motive}
In this section, we provide motivating examples in the application domain that fall into our problem formulation.

\subsection{Covid-19 Pandemic Prevention}

As the Covid-19 disease spreads all around the world, prevention and detection policies are implemented in different countries to relieve the global situation. Specifically, consider a medical institution that takes care of Covid-19 patients. The manager of this institution is receiving observations representing the severeness of symptoms from the patients each day. According to these symptoms, the manager decides whether a more harmful new variant has infected this institution and whether to take actions such as using heavier masks and implementing quarantines to stop the spread of new variants.

\subsection{Misinformation Intervention on Social Networks}
On social networks such as Twitter and Weibo, millions of fake news posts are created every day by social bots, which affect the judgement of real users (people). Consider an administrator who detects misinformation on its social platform. It tries to set up different levels of labels and warnings for potential fake news. Once a piece of information is retweeted, the administrator receives a new observation on the features of the retweeting user. Based on these features, the administrator decides whether the potential misinformation has contaminated real user communities, and which intervention action to implement to prevent users from retweeting it.

\subsection{Anomaly Detection in Communication Networks}
Consider a scenario where a communication link is provided and maintained by a service provider to multiple users or devices for transmitting data packets with a designated distribution of rates. Some user may become malicious at certain points and start to send packets with an higher rate for a long period time. Under such circumstances, the service provider may want to detect such behaviors and then take multiple interventions to throttle the data rate for some users.

\section{Model}\label{sec:model}
We first introduce the notations throughout our paper. Then, we define the diffusion model with a change-point that represents a mutation and the intervention model to control the mutated diffusion process with some assumptions. Finally, we present the problem formulation.

\subsection{Notation}
Throughout the paper, we use $1_{\{\cdot\}}$ to denote the indicator function. We use $\mathrm{Pr}$ to denote the probability measure and $\mathbb{E}_{X}$ to denote expectation with respect to random variable $X$. If the subscript is omitted, the expectation is taken with respect to all randomness. Vectors and matrices are written in boldface $\boldsymbol{a}$ and $\boldsymbol{A}$, with $a_i$ or $a(i)$ indicating the $i$-th entry of $\boldsymbol{a}$ and $A_{i,j}$ indicating the $(i,j)$-th element of $\boldsymbol{A}$.

\subsection{Sequential Diffusion Model}
\begin{figure}
	\centering
	\includegraphics[width=0.8\linewidth]{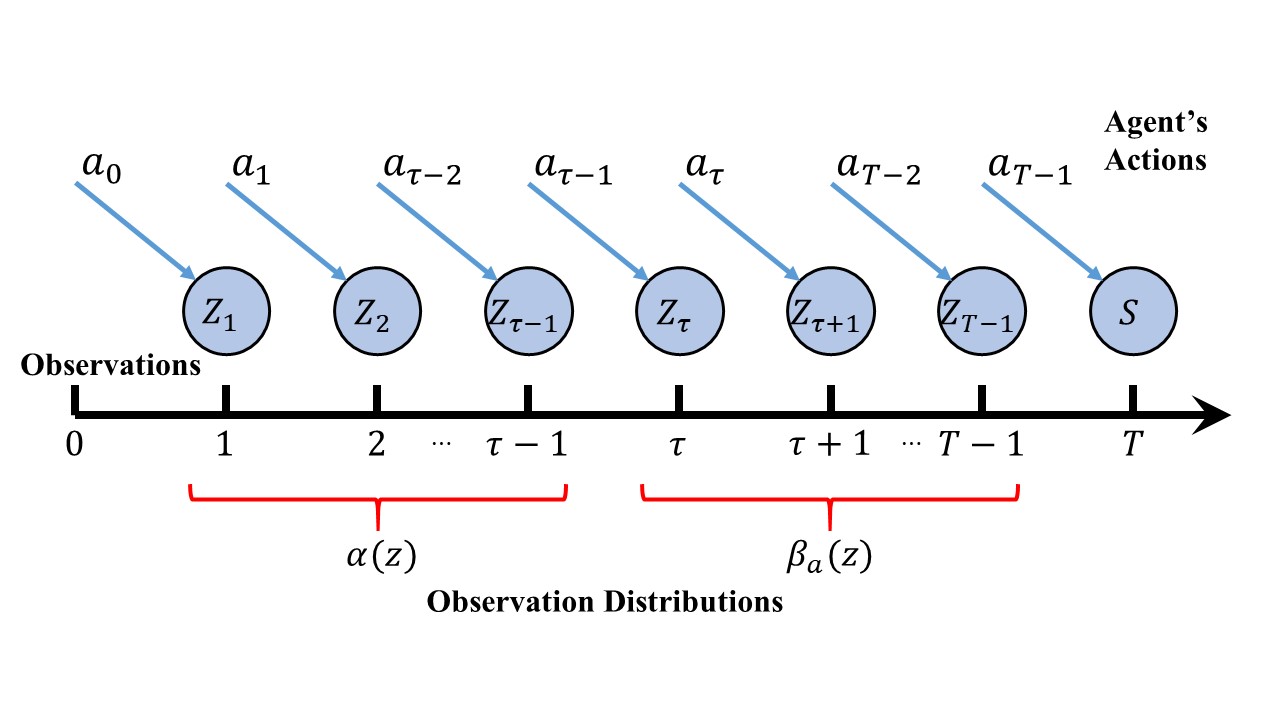}
	\caption{Illustration of the sequential diffusion model. At each time step, the agent receives an observation $z_t$ (blue circle), and then it makes an intervention $a_t$ to affect the observation $z_{t+1}$ in the next time step.}
	\label{fig:SeqModel}
\end{figure}

We consider a discrete-time process as shown in Fig.~\ref{fig:SeqModel} with a change-point $\tau$. An agent is observing this process and needs to intervene after a harmful mutation occurs at the change-point $\tau$. Let $t\in\{1,2,\cdots,T\}$ be the index of time slots, where the time horizon $T$ is a random variable following a geometric distribution with parameter $1-\rho$, i.e.,
\begin{align}
	\mathrm{Pr}(T=k)=\rho^{k-1}(1-\rho).
\end{align}
We assume the agent does not know $T$ a priori, but is aware of whether the process has terminated or not (e.g. a cure has been discovered or the disease suddenly disappears). For simplicity, we use $s_t={1}_{t\geq T}\in\{0,1\}$ to indicate whether the process has ended, i.e., $s_t=1$ means it has ended and $s_t=0$ means otherwise.

Let $\tau \in \{1,2,\cdots,T\}$ denote the change-point which is also unknown to the agent throughout the process. The change-point $\tau$ divides the time horizon into two phases: the pre-change phase before $\tau$ and the post-change phase after $\tau$. We assume that the change-point $\tau$ follows a geometric prior distribution with parameter $\lambda$:
\begin{align}
	\mathrm{Pr}(\tau=k)=\lambda(1-\lambda)^{k-1}.
\end{align}
For simplicity, we also denote $\theta_t={1}_{t\geq \tau}\in\{0,1\}$ to be the indicator whether the change point has happened.

At each time slot $t$, the agent receives an observation $z_t\in\{1,2,\cdots,Z\}$, which will be used by the agent to determine whether the change-point has occurred and how to intervene. We assume in the absence of any intervention actions, $z_t$ follows an i.i.d. distribution $\alpha(z)$ in the pre-change phase, and follows another i.i.d. distribution $\beta_0(z)$ in the post-change phase: 
\begin{subequations}
	\begin{align}
		\mathrm{Pr}(z_t=z|\theta_t=0)&=\alpha(z),\\
		\mathrm{Pr}(z_t=z|\theta_t=1)&=\beta_0(z).
	\end{align}
\end{subequations}

We remark that in this paper, we consider a harmful mutation occurs at the change-point $\tau$ and turns the original benign process into a harmful one. The terminal time $T$ represents the time at which either a cure is discovered or the mutation disappears. Considering the Covid-19 pandemic, the observations $z_t$ could be viewed as the symptoms of the $t$-th patient. A higher value of $z_t$ represents a severer symptom. For example, $z_t=1$ represents an asymptomatic patient while $z_t=Z$ represents hospitalization. 

\subsection{Sequential Intervention Model}

While receiving observations, the agent needs to decide whether to intervene and what actions to take. We assume the agent can take $A$ different actions representing $A$ different intervention levels, i.e., $a_t\in\{1,\cdots,A\}$, or it can choose to idle with $a_t=0$. We further assume that the intervention is only effective on the already mutated diffusion process, i.e., when $\theta_t=1$. The intervention will shift the original probability distribution $\beta_0(z)$ to a new distribution $\beta_a(z)$ according to the action $a$ which the agent chooses. Therefore, depending on whether the mutation has occurred, the observation $z_t$ is an i.i.d.\ sample from one of the following distributions:
\begin{subequations}\label{eq:obsdistribution}
	\begin{align}
		\mathrm{Pr}(z_t|\theta_t=0,a_{t-1}=a)&=\alpha(z_t),\quad \forall a\in\{0,\cdots,A\},\\
		\mathrm{Pr}(z_t|\theta_t=1,a_{t-1}=a)&=\beta_a(z_t), \quad \forall a\in\{0,\cdots,A\}.
	\end{align}
\end{subequations}

We remark that larger $a_t$ values indicate stricter intervention levels. Considering controlling Covid-19 pandemic, $a_t=1$ may represent voluntary social distancing and $a_t=A$ may mean strict lock-down. Throughout the paper, we assume that with the highest intervention level, we can suppress the mutated diffusion process to be as the original one, i.e.,  $\boldsymbol{\beta}_A=\boldsymbol{\alpha}$. Moreover, we assume that the agent can only gradually increase the intervention level $a_t$ and cannot decrease it. In other words, the intervention cannot be relaxed until the diffusion ends. We further impose the condition that the agent can only increase the intervention level by one at each time step, i.e., $a_{t-1}\leq a_t\leq a_{t-1}+1$. This is a technical assumption and does not affect the generality of our solution because we can choose the duration of each time slot to be sufficiently small so that the intervention level can go up multiple levels in a short period of time. We will show in experiment that leveraging this assumption doesn't harm our results empirically too much.

When controlling the diffusion process, the agent needs to consider the cost of an intervention action (e.g. the economic cost of lockdown) and the cost of the diffusion propagation (e.g. treatment of infected patients). We define $c_p^z$ to be the propagation cost associated with observation $z$, which can be viewed as the treatment of a patient with symptom $z$, and define $c_i^a$ to be the intervention cost of taking action $a$ at any time $t$. We assume that $c_i^a$ and $c_p^z$ are monotonically non-decreasing with $a$ and $z$ respectively, since stricter actions require more resources and severer symptoms require more treatments. Especially, the intervention cost of idle equals zero, i.e., $c_i^0=0$.

\subsection{Dominance Assumptions}

In order to model the strictness of intervention actions, we assume the distribution vector $\boldsymbol{\beta}_a$ of observations $z_t$ follows first order stochastic dominance and monotone likelihood ratio (MLR) ordering assumptions defined below. 
\begin{definition}[Stochastic Dominance]
	For two probability mass functions $\beta(z)$ and $\tilde{\beta}(z)$ on the same probability space, we say  $\boldsymbol{\beta}$ stochastically dominates $\tilde{\boldsymbol{\beta}}$, denoted by $\boldsymbol{\beta}\geq_s\tilde{\boldsymbol{\beta}}$, if 
	\begin{align}
	    \sum_{z=j}^Z\beta(z)\geq \sum_{z=j}^Z\tilde{\beta}(z), \quad \forall j\in\{1,2,\cdots,Z\}.
	\end{align}
\end{definition}
\begin{definition}[MLR Ordering]
	Given two probability mass functions $\beta(z)$ and $\tilde{\beta}(z)$ on the same probability space, we say  $\boldsymbol{\beta}$ MLR dominates $\tilde{\boldsymbol{\beta}}$, denoted by $\boldsymbol{\beta}\geq_r\tilde{\boldsymbol{\beta}}$, if 
	\begin{align}
	    \frac{\beta(z)}{\tilde{\beta}(z)}\leq\frac{\beta(z')}{\tilde{\beta}(z')}\quad \forall z\leq z'.
	\end{align}
\end{definition}

The two assumptions $\boldsymbol{\beta}\geq_s\tilde{\boldsymbol{\beta}}$ and $\boldsymbol{\beta}\geq_r\tilde{\boldsymbol{\beta}}$ both imply that under distribution ${\boldsymbol{\beta}},$ it is more likely to receive a higher value observation. Moreover, MLR ordering $\boldsymbol{\beta}\geq_r\tilde{\boldsymbol{\beta}}$ is a stronger assumption which directly implies stochastic dominance $\boldsymbol{\beta}\geq_s\tilde{\boldsymbol{\beta}}$~\cite{krishnamurthy2016partially}. In this paper, we assume the distribution of a less strict action dominates the distribution of a stricter one, i.e., 
\begin{align}
	\beta_a(z)\leq_r\beta_{a-1}(z), \quad \forall a\in \{1,2,\cdots,A\}.
\end{align}
We also assume that in the post-change phase after time $\tau$, stricter actions achieve lower overall cost: \begin{align}\label{assump:severerbetter}
	\mathbb{E}[c_p^{z_t}+c_i^{a_t}|\theta_t=1]< \mathbb{E}[c_p^{z_t}+c_i^{a_t-1}|\theta_t=1],\quad \forall a_t\geq 1.
\end{align}

\subsection{Problem Formulation}

The objective of the agent is to minimize the overall cost including both the propagation and intervention costs, without knowing $\tau$ and $T$. It requires the agent to detect the change-point $\tau$ as soon as possible along with selecting the optimal intervention action $a_t$. Let $\mu$ be a policy which maps all past information $\mathcal{F}_t$ to the intervention action $a_t$. Denote by $\Pi_{NA}$ the set of all non-anticipating policies. The quickest intervention problem is defined below: 
\begin{pb}[Quickest Intervention]\label{pb:costmin}
	\begin{subequations}
		\begin{align}
			&\min_{\mu\in\Pi_{NA}} C^\mu =\mathbb{E}_\mu\left[c_i^{a_0}+\sum_{t=1}^{T-1}\left(c_p^{z_t}+c_i^{a_t}\right)\right],\\
			\text{s.t.}\quad &a_{t-1}\leq a_t\leq \min\{a_{t-1}+1,A\}, \quad \forall 1\leq t\leq {T-2}.\label{eq:consintervene}
		\end{align}
	\end{subequations}
\end{pb}

\section{A POMDP Approach to Quickest Intervention}\label{sec:resolve}

In this section, we view Problem \ref{pb:costmin} as a POMDP, and then convert it into a fully observed MDP by replacing the unobserved change-point indicator $\theta_t$ with the belief state $\pi_t$. With a grid approximation approach, we solve the optimal policy from the belief MDP. 

\subsection{Partially Observed MDP Formulation}
The quickest intervention Problem \ref{pb:costmin} can be formulated as a partially observed MDP $\mathcal{M}=(\mathcal{S},\mathcal{O},\mathcal{A},\boldsymbol{P}(a),\boldsymbol{O}(a),C(z,a))$ with the unknown indicator $\theta_t$ of the change-point $\tau$. In particular, the POMDP is defined as follows:
\begin{itemize}
	\item \textbf{State Space:} $S_t=(s_t,\theta_t,\tilde{a}_t)\in \{0,1\}^2\times\{0,1,\cdots,A\}$. Here, $s_t$ is the indicator whether the MDP has stopped; $\theta_t$ is the indicator whether the change point has happened; and $\tilde{a}_t$ is the current intervention level.
	\item \textbf{Observation Space:} $z_t\in\{1,\cdots,Z,S\}$. Here, $z_t$ is the observation that the agent receives at time $t$, which helps the agent decide intervention actions. Notice that $z_t=S$ means the MDP has stopped.
	\item \textbf{Action Space:} $a_t\in\{0,1,\cdots,A\}$. Here, action $a_t$ is the interventions that the agent takes at time slot $t$ to control the mutated process.
	\item \textbf{State Transition Matrix:} At each time step, the change-point occurs with probability $\lambda$ and the MDP stops with probability $1-\rho$, i.e., 
	\[
	\boldsymbol{P}_{s_t}=
	\left[
	\begin{matrix}
		\rho & 1-\rho \\
		0 & 1\\
	\end{matrix}
	\right],
	\boldsymbol{P}_{\theta_t}=
	\left[
	\begin{matrix}
		1-\lambda & \lambda \\
		0 & 1\\
	\end{matrix}
	\right], \tilde{a}_t=a_{t-1}.
	\]
	\item \textbf{Observation Model:} At each slot $t$, the agent observes whether the process has ended. If not, it receives an observation $z_t$ following the distribution in Eq.~\eqref{eq:obsdistribution}.
	\item \textbf{One-step Cost:} With observation $z_t$ and action $a_t,$ the agent incurs a cost of $C(z_t, a_t)=c_p^{z_t}+c_i^{a_t}$. 
\end{itemize}

It is well-known in \cite{krishnamurthy2016partially} that solving a general POMDP is difficult. In the next subsection, we replace the unknown state dimension $\theta_t$ with its posterior $\pi_t$ given all past information. Then, we convert the partially observed MDP into a fully observed MDP which is also called belief MDP.

\subsection{Belief MDP Formulation}
To convert a POMDP to an MDP, we replace $\theta_t$ with belief $\pi_t=\mathrm{Pr}(\theta_t=1|\mathcal{F}_t)$, where $\mathcal{F}_t$ is the $\sigma$-algebra that contains all past observations and actions, i.e., $\mathcal{F}_t=\{a_0,z_1,a_1,\cdots,a_{t-1},z_t\}$. Suppose at time $t$, the posterior of $\theta_t=1$ is $\pi_t$. Then, at the beginning of time slot $t+1$, before receiving observation $z_{t+1}$, the posterior changes to $\tilde{\pi}_{t+1}$  as follows, due to the geometric prior distribution,
\begin{align}\label{eq:posterior1}
	\tilde{\pi}_{t+1}=&\nonumber\mathrm{Pr}(\theta_{t+1}=1|\mathcal{F}_t,a_t)\\
	\overset{(a)}{=}&\nonumber\mathrm{Pr}(\theta_{t}=1|\mathcal{F}_t)\mathrm{Pr}(\theta_{t+1}=1|\theta_{t}=1,\mathcal{F}_t)+\mathrm{Pr}(\theta_{t}=0|\mathcal{F}_t)\mathrm{Pr}(\theta_{t+1}=1|\theta_{t}=0,\mathcal{F}_t)\\
	\overset{(b)}{=}&\pi_t+\lambda(1-\pi_t),
\end{align}
where equality ($a$) holds because the change-point is independent of intervention $a_t,$ and equality ($b$) is obtained from the fact that the evolution of $\theta_{t}$ is a time-homogeneous Markov chain with transition matrix $\boldsymbol{P}_{\theta_{t}}$. Therefore, adding the observation $z_{t+1}$ and action $a_t$ into the $\sigma$-algebra, the update of the posterior $\pi_{t+1}$ is as follows: 
\begin{align}\label{eq:posterior2}
	\pi_{t+1}=&\nonumber\mathrm{Pr}(\theta_{t+1}=1|\mathcal{F}_{t+1})\\
	\overset{(a)}{=}&\nonumber\frac{\mathrm{Pr}(\theta_{t+1}=1|\mathcal{F}_{t},a_t)\mathrm{Pr}(z_{t+1}|\theta_{t+1}=1,\mathcal{F}_{t},a_t)}{\mathrm{Pr}(z_{t+1}|\mathcal{F}_t,a_t)}\\
	=&\frac{\tilde{\pi}_{t+1}\beta_{a_t}(z_{t+1})}{\mathrm{Pr}(z_{t+1}|\mathcal{F}_t,a_t)},
\end{align}
where equality ($a$) follows from Bayes' rule. Note that from the total probability theorem, we have:
\begin{align}\label{eq:posterior3}
	\mathrm{Pr}(z_{t+1}|\mathcal{F}_t,a_t)=&\nonumber\mathrm{Pr}(\theta_{t+1}=1|\mathcal{F}_{t},a_t)\mathrm{Pr}(z_{t+1}|\theta_{t+1}=1,\mathcal{F}_{t},a_t)\\
	&\nonumber+\mathrm{Pr}(\theta_{t+1}=0|\mathcal{F}_{t},a_t)\mathrm{Pr}(z_{t+1}|\theta_{t+1}=0,\mathcal{F}_{t},a_t)\\
	=&\tilde{\pi}_{t+1}\beta_{a_t}(z_{t+1})+(1-\tilde{\pi}_{t+1})\alpha(z_{t+1}).
\end{align}
Substituting Eq.~\eqref{eq:posterior1} and Eq.~\eqref{eq:posterior3} into Eq.~\eqref{eq:posterior2}, we conclude that the posterior can be updated iteratively as follows:
\begin{align}\label{eq:posteriorupdate}
	\pi_{t+1}=&T_{a_t}(\pi_t,z_{t+1})\nonumber\\
	=&\frac{\tilde{\pi}_{t+1}\beta_{a_t}(z_{t+1})}{\tilde{\pi}_{t+1}\beta_{a_t}(z_{t+1})+(1-\tilde{\pi}_{t+1})\alpha(z_{t+1})},
\end{align}
where $\tilde{\pi}_{t+1}=\pi_t+\lambda(1-\pi_t)$. The following theorem shows that the optimal policy for the quickest intervention problem can be obtained from studying the fully observed belief MDP.
\begin{theorem}\label{theo:beliefMDP}
Solving the quickest intervention problem defined on the POMDP with $S_t=(s_t,\theta_t,\tilde{a}_t)$ and unknown state $\{\theta_t\}$ is equivalent to solving the problem on a fully observed MDP with $S_t=(\pi_t,\tilde{a}_t,z_t)$ where $\pi_t$ replaces the unknown state $\theta_{t}$. In the optimal solution, the intervention action $a_t$ at slot $t$ depends purely on the belief state $\pi_t$.
\end{theorem}

The proof of Theorem~\ref{theo:beliefMDP} is provided in appendix. Based on the theorem above, the quickest intervention problem is converted to an MDP with the belief state $\pi_t$. To solve this MDP, we define $V^t(\pi,\tilde{a},z)$ to be the value function of state $S=(\pi,\tilde{a},z)$ at time $t$, i.e. 
\begin{align}
	V^t(\pi,\tilde{a},z)=&\min_{\mu\in \Pi_{NA}}\mathbb{E}_\mu\left[\left.\sum_{k=t}^TC(z_k,\mu(\pi_k))\right|S_t=(\pi,\tilde{a},z)\right].
\end{align}
We further define $J_a^t(\pi,\tilde{a},z)$ to be the action-value function of choosing $a_t=a$ at state $S=(\pi,\tilde{a},z)$ at time $t$ as follows. 
\begin{align}
	J_a^t(\pi,\tilde{a},z)=&C(z_t,a)
	+\min_{\mu}\mathbb{E}_\mu\left[\left.\sum_{k=t+1}^TC(z_k,\mu(\pi_k))\right|S_t=(\pi,\tilde{a},z),a_t=a\right].
\end{align}

For simplicity, in the following analysis, we set the value function $V^t(\pi,\tilde{a},z)$ to be zero for any state that the MDP has stopped with stopping state $z=S$, i.e., $V^t(\pi,\tilde{a},S)=0$ for any $\pi$, $\tilde{a}$ and $t$. Also, we define $\sigma_a(\pi,z)$ to be the distribution of observation $z_{t+1}$ given posterior $\pi_t$ and action $a_t$ as:
\begin{align}
\sigma_a(\pi,z)=&\mathrm{Pr}(z_{t+1}=z|\pi_t=\pi,a_t=a)\nonumber\\
=&\alpha(z)(1-\tilde{\pi})+\beta_{a}(z)\tilde{\pi},
\end{align}
where $\tilde{\pi}=\pi+\lambda(1-\pi)$ is the sudo-posterior before receiving the next observation $z_{t+1}$. 

Recall that $\rho$ is the probability the diffusion process continues. Therefore, the Bellman equation for action-value function $J_a^t(\pi,\tilde{a},z)$ of the belief state MDP can be recursively expressed as follows:
\begin{align}\label{eq:bellmanoriginal1}
	J_a^t(\pi,\tilde{a},z)
	=&\nonumber c_i^a+c_p^z+(1-\rho)V^{t+1}(\pi,a,S)
	+\rho\sum_{z'=1}^Z\sigma_a(\pi,z')V^{t+1}(T_a(\pi,z'),a,z')\\
	=&c_i^a+c_p^z+\rho\sum_{z'=1}^Z\sigma_a(\pi,z')V^{t+1}(T_a(\pi,z'),a,z').
\end{align}

Note that according to Eq.~\eqref{eq:consintervene}, the agent can only choose $a_t$ between $\tilde{a}_t$ and $\min\{\tilde{a}_t+1,A\}$. The value function can also be expressed recursively as:
\begin{align}\label{eq:bellmanoriginal2}
	V^t(\pi,\tilde{a},z)=\min\{J_{\tilde{a}}^t(\pi,\tilde{a},z),J_{\min\{\tilde{a}+1,A\}}^t(\pi,\tilde{a},z)\}.
\end{align}

Since this MDP is a stochastic shortest path problem where the optimal policy is time-homogeneous and does not depend on time $t$~\cite{BertsekasDP}, we omit the superscript $t$ from now on. Notice that from Eq. \eqref{eq:bellmanoriginal1}, $J_a(\pi,\tilde{a},z)$ is independent of $\tilde{a}$, and state $z$ is not involved in deciding the optimal action. We can subtract the propagation cost $c_p^z$ in the action-value function, and then remove state $z$ and $\tilde{a}$ in $J_a^t(\pi,\tilde{a},z)$. More specifically, denote $$J_a(\pi)=J_a^t(\pi,\tilde{a},z)-c_p^z$$ to be a modified action-value function, and denote $V_{\tilde{a}}(\pi)$ to be the modified value function related to $J_a(\pi)$. Since they do not depend on $z$, we will use $z$ to replace $z'$ in the remaining paper. Therefore, we can simplify the Bellman equation \eqref{eq:bellmanoriginal1} and \eqref{eq:bellmanoriginal2} as follows:
\begin{subequations}\label{eq:bellman}
    \begin{align}
	J_a(\pi)=&c_i^a+\rho\sum_{z=1}^Z\sigma_a(\pi,z)\left[c_p^{z}+V_a(T_a(\pi,z))\right],\label{eq:bellman1}\\
	V_{\tilde{a}}(\pi)=&\min\{J_{\tilde{a}}(\pi),J_{\min\{{\tilde{a}}+1,A\}}(\pi)\}.\label{eq:bellman2}
\end{align}
\end{subequations}

By solving the above Bellman equations, the agent obtains an optimal policy $\mu^*(\pi)$ for the quickest intervention problem defined in Problem \ref{pb:costmin}. In the next subsection, we will prove the concavity and monotonicity of value functions. These properties will be useful in deriving a low-complexity algorithm in later sections.

\subsection{A Structural Result for the Optimal Policy}
We present an important property that the value functions of the belief MDP possess concavity and monotonicity. 
\begin{lemma}\label{lemma:concavity}
	For any action $a$ and current intervention level $\tilde{a}$, the value function $V_{\tilde{a}}(\pi)$ and the action-value function $J_a(\pi)$ are both concave and monotonically non-decreasing in belief state $\pi$. Furthermore, $V_A(\pi)$ and $J_A(\pi)$ are constant: 
	\begin{align}
		V_A(\pi)=J_A(\pi)=\frac{c_i^A+\rho\sum_{z=1}^Z\alpha(z)c_p^z}{1-\rho}.
	\end{align}
\end{lemma}

The proof of Lemma~\ref{lemma:concavity} is provided in appendix. However, since $\pi$ is continuous, directly solving the Bellman equations is difficult. In the next subsection, we present a grid approximation approach to obtain a nearly optimal intervention policy.

\subsection{Solving Belief MDP through Grid Approximation}
\begin{algorithm}[h]
	\caption{Grid Approximation Algorithm for $\mu^*(\pi)$}\label{algo:grid}
	\begin{algorithmic}[1]
		\REQUIRE grids $\{B_j\}_{1\leq j\leq N}$, representatives $\{\pi_{B_j}\}_{1\leq j\leq N}$, error tolerance $\epsilon$.
		\STATE \textbf{Initialization:} $k=0$, $V_{\tilde{a}}^{(0)}(\pi_{B_j})=0$, $J_a^{(0)}(\pi_{B_j})=0$.
		\WHILE{$\exists B_j$ s.t. $|V_{\tilde{a}}^{(k)}(\pi_{B_j})-V_{\tilde{a}}^{(k-1)}(\pi_{B_j})|>\epsilon$,}
		\FOR{$j=1:N$}
		\FOR{$a=0:A$}			
		\STATE Update the action-value function $J_a^{(k+1)}(\pi_{B_j})$ according to $V_{\tilde{a}}^{(k)}(\pi_{B_j})$ and Eq.~\eqref{eq:bellman1}.
		\ENDFOR
		\ENDFOR
		\FOR{$j=1:N$}
		\FOR{${\tilde{a}}=0:A$}
		\STATE Update the value function $V_{\tilde{a}}^{(k+1)}(\pi_{B_j})$ and action $\mu(\pi_{B_j})$ according to $J_a^{(k+1)}(\pi_{B_j})$ and Eq.~\eqref{eq:bellman2}.
		\ENDFOR
		\ENDFOR
		\ENDWHILE
		\FOR{$j=1:N$}
		\STATE Set action $\mu(\pi)$ for any $\pi\in B_j$ to be $\mu(\pi_{B_j})$.
		\ENDFOR
		\ENSURE Nealy optimal policy $\{\mu(\pi)\}$.
	\end{algorithmic}
\end{algorithm}

In this section, we provide in Algorithm~\ref{algo:grid} a grid approximation approach to solve the belief MDP~\cite{Lovejoygrid,VVVbinning}. We first divide $\pi$ into properly designed grids $\{B_1,B_2,\cdots,B_N\}$, where $N$ is a proper number. One such design of grids may be uniform grids. Note that $B_i$'s are non-intersected sections of real values satisfying $\cup_{j=1}^N B_j=[0,1]$. Then, we select one element $\pi_{B_i}$ in each grid $B_j$ as its representative. The rest follows the standard value iteration~\cite{BertsekasDP}. 

Notice that this approach provides nearly optimal policies when the number of grids is large so that all grids are sufficiently small. However, since the performance of the obtained policy depends on the number of grids, the computational complexity is still very high when $N$ is large. In the next section, we exploit linear approximation to obtain a low-complexity algorithm which has a nearly optimal provable guarantee in the \emph{``local intervention''} regime.

\section{A Low-Complexity Algorithm with Nearly Provable Guarantee}\label{sec:small}

In this section, we address the computational complexity of grid approximation by providing a heuristic low-complexity algorithm based on an approximated Bellman equation. We introduce the \emph{``local intervention''} regime where the value function can be approximated accurately with the first-order Taylor's approximation. In this regime, we prove the nearly submodularity of the value function. Then, we design a threshold-based intervention policy, where the thresholds are easily calculable with closed-form expression.

\subsection{``Local Intervention'' Regime Analysis}

To derive the low-complexity algorithm based on the Taylor's approximation, we consider the problem in the \emph{``local intervention''} regime where all consecutive intervention levels are close to each other, and thus also close to the benign distribution. The formal definition of \emph{``local intervention''} regime is presented in Assumption~\ref{assump:small}. The definition says that for a sufficiently small constant $\delta$ and any arbitrary discounted factor $\gamma\in(0,1)$, the differences in cumulative distribution functions of $\boldsymbol{\beta}_a$'s for any state $z$ are all very small.
\begin{assumption}(\emph{``Local Intervention''})\label{assump:small}
	For any action $a\geq 1$, the observation distribution $\beta_a(z)$ follows:
	\begin{align}
		\gamma\delta\leq\left|\sum_{z=j}^Z\beta_{a-1}(z)-\sum_{z=j}^Z\beta_{a}(z)\right|\leq \delta, \quad \forall j\geq 2,
	\end{align}
	where $\delta$ is a sufficiently small real value and $\gamma\in(0,1)$.
\end{assumption} 

This assumption implies that
$\beta_{a}(z)-\alpha(z)=\mathcal{O}(\delta)$ for any $a$. In later analysis, we will include superscript $\delta$ on any value that is related to $\delta$. For the simplicity of notation, we define the following three values:
\begin{align}
    A_p^a=&\sum_{z=1}^Z(\beta_a(z)-\alpha(z))c_p^z,\\
	D_{p}^a=&\sum_{z=1}^Z(\beta_{a}(z)-\beta_{a-1}(z))c_p^z,\\
	D_{i}^a=&c_i^a-c_i^{a-1}.
\end{align}
It is obvious to notice that $D_p^a=A_p^a-A_p^{a-1}$. According to the \emph{``local intervention''} assumption, we have $A_p^a=\Theta(\delta)$ and $D_p^a=\Theta(\delta)$. 

We then analyze the posterior update function Eq.~\eqref{eq:posteriorupdate} with Taylor's expansion evaluated at $\alpha(z)$ when $\delta$ is small enough:
\begin{align}
	T_a^\delta(\pi,z)=&\nonumber\frac{\tilde{\pi}\beta_a(z)}{\tilde{\pi}\beta_a(z)+(1-\tilde{\pi})\alpha(z)}\\
	=&\nonumber \frac{\tilde{\pi}(\alpha(z)+\beta_{a}(z)-\alpha(z))}{\tilde{\pi}(\alpha(z)+\beta_{a}(z)-\alpha(z))+(1-\tilde{\pi})\alpha(z)}\\
	=& \tilde{\pi} + \frac{(1-\tilde{\pi})\tilde{\pi}}{\alpha(z)}(\beta_{a}(z)-\alpha(z))+\mathcal{O}(\delta^2)\label{eq:posteriorupdatesmall},
\end{align}
where $\tilde{\pi}=\pi+\lambda(1-\pi)$.

Substituting Eq.~\eqref{eq:posteriorupdatesmall} into the expression of $J_a(\pi)$ in Eq.~\eqref{eq:bellman1},  we obtain:
\begin{align}
	J^\delta_a(\pi)=&\nonumber c_i^a+\rho\sum_{z=1}^Z\left[\alpha(z)+\tilde{\pi}(\beta_{a}(z)-\alpha(z))\right]
	\cdot\left[c_p^z+ V_a\left(\tilde{\pi} + \frac{(1-\tilde{\pi})\tilde{\pi}}{\alpha(z)}(\beta_{a}(z)-\alpha(z))+\mathcal{O}(\delta^2)\right)\right]\\
	\overset{(a)}{=}&\left[c_i^a+\rho\sum_{z=1}^Z\alpha(z)c_p^z\right]+\tilde{\pi}\rho A_p^a+\rho V_a(\tilde{\pi})+\mathcal{O}(\delta^2),\label{eq:bellman1small}
\end{align}
where equality $(a)$ is obtained by Taylor's expansion of $V_a(\cdot)$ evaluated at point $\tilde{\pi}$, as well as the definition that $\sum_{z=1}^Z(\beta_a(z)-\alpha(z))c_p^z=A_p^a$. Also note that the first-order terms cancels due to $\sum_{z=1}^Z\alpha(z)=\sum_{z=1}^Z\beta_a(z)=1$. Notice that the first term $[c_i^a+\rho\sum_{z=1}^Z\alpha(z)c_p^z]$ in Eq.~\eqref{eq:bellman1small} is independent of $\pi$, and the second term $\tilde{\pi}\rho A_p^a$ is linear with $\pi$. In order to approximate the value function $J_a^\delta(\pi)$ linearly, we need to focus on the non-linear terms $\rho V_a(\tilde{\pi})$ and $\mathcal{O}(\delta^2)$. Before we proceed, we first present the definitions of submodular functions and nearly-submodular functions.

\begin{definition}\label{def:submodular}
	An action-value function $J_a(\pi)$ is called submodular in pair $(a,\pi)$ if it satisfies $$\Delta_a(\pi)= J_{a+1}(\pi)-J_a(\pi)$$ is non-increasing with belief $\pi$.
\end{definition}

From Topkis' lemma ~\cite[Theorem 4.1]{Topkis78submodular}, if an action-value function $J_a(\pi)$ is submodular, the solution to its ``argmin'' problem, i.e., $$a^*(\pi)=\arg\min_a J_a(\pi),$$ is monotonically non-decreasing. Hence, a threshold exists since the action space is finite. 

Next, we check the submodularity of our value function. For a given action $a<A$, subtract Eq.~\eqref{eq:bellman1small} for consecutive intervention actions and we get:
\begin{align}\label{eq:smalldelta}
	\Delta^\delta_a(\pi)=&\nonumber
	J_{a+1}^\delta(\pi)-J_a^\delta(\pi)\\
	=&\nonumber \left[c_i^{a+1}-c_i^{a}\right]+\tilde{\pi}\left[\rho\sum_{z=1}^Z(\beta_{a+1}(z)-\beta_a(z))c_p^z\right]+\rho\left[V_{a+1}^\delta(\tilde{\pi})-V_{a}^\delta(\tilde{\pi})\right]+\mathcal{O}(\delta^2)\\
	=& D_i^{a+1}+\tilde{\pi}\rho D_p^{a+1}+\rho\left[V_{a+1}^\delta(\tilde{\pi})-V_{a}^\delta(\tilde{\pi})\right]+\mathcal{O}(\delta^2).
\end{align}

Recall that in the \emph{``local intervention''} regime, $\rho D_p^{a+1}=\Theta(\delta)$ is the same order as $\delta$. Heuristically as $\delta$ is small enough, the higher order term $\mathcal{O}(\delta^2)$ can be neglected. Therefore, we can accurately approximate the difference $\Delta_a^\delta(\pi)$ in the \emph{``local intervention''} regime through an approximated difference function $\tilde{\Delta}_a^\delta(\pi)$ defined as follows:
\begin{align}\label{eq:bellmanapproxdiff}
    \tilde{\Delta}_a^\delta(\pi)=D_i^{a+1}+\tilde{\pi}\rho D_p^{a+1}+\rho\left[\tilde{V}_{a+1}^\delta(\tilde{\pi})-\tilde{V}_{a}^\delta(\tilde{\pi})\right],
\end{align}
where $\tilde{V}_a(\pi)$ and $\tilde{J}_a(\pi)$ are defined similarly by neglecting the higher orders in the original expression Eq.~\eqref{eq:bellman1small} as follows:
\begin{subequations}\label{eq:bellmanapprox}
\begin{align}
    \tilde{J}_a(\pi)=&\left[c_i^a+\rho\sum_{z=1}^Z\alpha(z)c_p^z\right]+\tilde{\pi}\rho A_p^a+\rho \tilde{V}_a(\tilde{\pi}),\label{eq:bellmanapprox1}\\
    \tilde{V}_a(\pi)=&\min\{\tilde{J}_a(\pi),\tilde{J}_{\min\{a+1,A\}}(\pi)\}.\label{eq:bellmanapprox2}
\end{align}

\end{subequations}
Eq.~\eqref{eq:bellmanapprox} can be interpreted as a mean approximation to the original Bellman equation in Eq.~\eqref{eq:bellman}. Instead of computing the next-stage belief $T_a(\pi,z)$ and evaluating the corresponding value function $V_a(T_a(\pi,z))$ for each observation $z$ as in Eq.~\eqref{eq:bellman1}, we adopt the expected average belief update $\tilde{\pi}$ and substitute
$$\sum_{z=1}^Z\sigma_a(\pi,z)V_a\left(T_a(\pi,z)\right)$$
with
$$V_a\left(\sum_{z=1}^Z\sigma_a(\pi,z)T_a(\pi,z)\right)=V_a(\tilde{\pi}).$$

With the approximated MDP defined in Eq.~\eqref{eq:bellmanapprox} and similar to submodularity, we introduce the definition of nearly-submodular functions in the \emph{``local intervention''} regime.
\begin{definition}\label{def:nearlysubmodular}
	We define an action-value function $J^\delta_a(\pi)$ to be nearly-submodular in pair $(a,\pi)$ in the \emph{``local intervention''} regime with $\delta$, if its approximated difference function $$\tilde{\Delta}_a^\delta(\pi)= \tilde{J}_{a+1}^\delta(\pi)-\tilde{J}_a^\delta(\pi)$$ defined in Eq.~\eqref{eq:bellmanapproxdiff} is monotonically non-decreasing of $\pi$.
\end{definition}

From the expression of $\tilde{\Delta}_a^\delta(\pi)$ above in Eq.~\eqref{eq:bellmanapproxdiff}, we prove the nearly-submodularity of $J_a^\delta(\pi)$ through induction in the following Theorem~\ref{theo:thressmall}.
\begin{theorem}\label{theo:thressmall}
	The belief MDP and its approximation with posterior $\pi$ indicating whether change-point $\tau$ has occurred possess the following properties in the \emph{``local intervention''} regime with $\delta$:
	\begin{itemize}
		\item The action-value function $J_a^\delta(\pi)$ is nearly-submodular with pair $(a,\pi)$.
		\item The policy $\tilde{\mu}^{*,\delta}(\pi)$ that solves the approximated Bellman equation \eqref{eq:bellmanapprox} is monotonically non-decreasing, and thus possesses a threshold structure. That is, there exists a set of non-decreasing thresholds denoted as $\{\tilde{\pi}_a^{*,\delta}\}_{1\leq a\leq A}$, such that policy $\tilde{\mu}^{*,\delta}(\pi)$ switches to an action no less than $a$ iff $\pi$ exceeds threshold $\tilde{\pi}_a^{*,\delta}$.
	\end{itemize}
\end{theorem}

The proof of Theorem~\ref{theo:thressmall} is provided in appendix. From intuition, we expect that in the \emph{``local intervention''} regime, the optimal policy $\mu^{*,\delta}(\pi)$ to the belief MDP behaves like a threshold-based policy just as $\tilde{\mu}^{*,\delta}(\pi)$, and the thresholds for the two policies will be very close to each other. This property will be identified in our simulations.

Recall that the optimal cost $C^{\mu^*}$ of Problem \ref{pb:costmin} can be calculated from value function $V_0(0)$ defined in Eq.~\eqref{eq:bellman} with state $\tilde{a}=0$ and $\pi=0$, i.e., $C^{\mu^*}=V_0(0)$. Similarly, we define an approximated optimal cost $\tilde{C}$ for the approximated MDP in Eq.~\eqref{eq:bellmanapprox} as follows:
\begin{align}
    \tilde{C}=\tilde{V}_0(0).
\end{align}
Intuitively, $\tilde{C}$ will also be close to $C^{\mu^*}$. The following corollary provides a closed-form expression on $\tilde{C}$.

\begin{corollary}\label{cor:costbound}
	If all the thresholds $\{\tilde{\pi}_a^{*,\delta}\}_{1\leq a\leq A}$ are strictly within section $[0,1)$, the total cost $\tilde{C}$ obtained by solving Bellman equation \eqref{eq:bellmanapprox} is computed as follows:
	\begin{align}
		\tilde{C}
		= &\sum_{a=1}^A\frac{\rho^{t_a}}{1-\rho}D_i^a+\sum_{a=1}^A\left(\frac{\rho^{t_a+1}}{1-\rho}-\frac{[\rho(1-\lambda)]^{t_a+1}}{1-\rho(1-\lambda)}\right)D_p^a 
		+\frac{\rho\sum_{z=1}^Z\alpha(z)c_p^z}{1-\rho}+\frac{\lambda\rho A_p^0}{(1-\rho)[1-\rho(1-\lambda)]}.
	\end{align}
	where $t_a=\left\lceil\frac{\ln(1-\tilde{\pi}_a^{*,\delta})}{\ln(1-\lambda)}\right\rceil$ is the number of steps switching from action $0$ to an action no less than $a$.
\end{corollary}

The proof of Corollary~\ref{cor:costbound} is provided in appendix. However, calculating the thresholds $\{\tilde{\pi}_a^{*,\delta}\}_{1\leq a\leq A}$ is still computationally expensive. Therefore, we provide bounds on these thresholds in Corollary~\ref{cor:thresboundsmall}, which have close-form expressions. 
\begin{corollary}\label{cor:thresboundsmall}
	The thresholds $\{\tilde{\pi}_a^{*,\delta}\}_{1\leq a\leq A}$ can be upper and lower bounded with closed form expressions. Specifically, for a single threshold $\tilde{\pi}_a^{*,\delta}$ that $\tilde{\mu}^{*,\delta}(\pi)$ switches from $a-1$ to $a$, the following holds:
	\begin{align}
		\bar{\pi}_a+\frac{\sum_{j=a+1}^{A}\rho^{j-a}\left(D_i^j+\rho D_p^j\right)}{-(1-\lambda)\rho D_p^a}\leq
		\tilde{\pi}_a^{*,\delta}
		\leq \bar{\pi}_a,
	\end{align}
where the upper bound threshold $\bar{\pi}_a$ is
\begin{align}
	\bar{\pi}_a=-\frac{D_i^a}{(1-\lambda)\rho D_p^a}-\frac{\lambda}{1-\lambda}. 
\end{align}
\end{corollary}

The proof of Corollary~\ref{cor:thresboundsmall} is provided in appendix. Heuristically, from Corollary~\ref{cor:thresboundsmall}, if the optimal policy $\mu^{*,\delta}(\pi)$ also possesses a threshold $\hat{\pi}_a^{*,\delta}$, it is highly likely that $\hat{\pi}_a^{*,\delta}$ lies in the same section or its $\mathcal{O}(\delta)$ neighborhood when $\delta$ is small. On the other hand, the upper bound $\bar{\pi}_a$ serves as an approximated threshold which is easily calculable. In the next section, we present a low-complexity threshold-structured policy based on the upper bounds $\{\bar{\pi}_a\}_{1\leq a \leq A}$.

\subsection{Low-complexity Policy based on Upper Bounds}
\begin{algorithm}[h]
	\caption{Low-complexity Policy}\label{algo:lowcomp}
	\begin{algorithmic}[1]
		\STATE Initialize the thresholds to be $\bar{\pi}_{A+1}\leftarrow1$ and $\bar{\pi}_a\leftarrow\min\left\{-\frac{D_i^a}{(1-\lambda)\rho D_p^a}-\frac{\lambda}{1-\lambda},\bar{\pi}_{a+1}\right\}$ for $a\leq A$.
		\WHILE{MDP doesn't stop}
			\STATE Obtain observation $z_t$.
			\STATE Update belief $\pi_t=T(\pi_{t-1},a_{t-1},z_t)$.
			\STATE Implement action $a_t=a$, s.t. $\bar{\pi}_a \leq \pi_t \leq\bar{\pi}_{a+1} $.
		\ENDWHILE
	\end{algorithmic}
\end{algorithm}

In this section, we propose our low-complexity policy $\bar{\mu}$. The policy and its implementation are summarized in Algorithm~\ref{algo:lowcomp}. The intuition is simple: since the upper bound $\bar{\pi}_a$ is very close to the original threshold $\tilde{\pi}_a^{*,\delta}$, choosing $\bar{\pi}_a$ to be the threshold is actually a conservative policy. In Sec~\ref{sec:simu}, we show from the numerical simulations that the low-complexity policy's performance is nearly the same as the optimal one even beyond the \emph{``local intervention''} regime.

\section{Simulations}\label{sec:simu}

In this section, we first provide a lower bound on the total cost $C^{\mu^*}$ along with the definition of cost regret. These concepts are very important in our simulations. Then, we will proceed to present the numerical simulation results.

\subsection{Lower Bound and Regret}\label{sec:lb}
Consider an oracle that knows the time horizon $T$ and the change-point $\tau$ in advance. Then after relaxing the incremental constraint in Eq.~\eqref{eq:consintervene}, the oracle can minimize the overall cost with a very simple oracle policy $\mu_o$: choose $a_t=A$ when $\tau\leq t <T-1$ and $a_t=0$ otherwise. The ``stricter is better'' assumption in Eq.~\eqref{assump:severerbetter} ensures its optimality. We derive the cost of the oracle policy in the following Theorem~\ref{theo:oracle}.
\begin{theorem}\label{theo:oracle}
	If the change point $\tau$ and horizon $T$ is known ahead of time, then the optimal oracle action $a_t$ is as follows:
	\begin{itemize}
		\item When $0\leq t\leq \tau-2$, $a_t=0$.
		\item When $\tau-1 \leq t\leq T-2 $, $a_t=A$.
		\item When $t=T-1$, $a_{t}=0$.
	\end{itemize}
	The expected cost performance of oracle policy $\mu_o$ is
	\begin{align}
		C^{\mu_o}=\frac{\rho\sum_{z=1}^Z\alpha(z)c_p^z}{1-\rho}+c_i^A\left[\frac{\rho}{1-\rho}-\frac{\rho(1-\lambda)}{1-\rho(1-\lambda)}\right].
	\end{align}
\end{theorem}

The proof of Theorem~\ref{theo:oracle} is provided in appendix. $C^{\mu_o}$ is the lower bound on the expected total cost, which only relates to system parameters such as $\rho$, $\lambda$, and $\boldsymbol{\alpha}$. To better understand the performance of our proposed policy, we study the cost regret for a given policy $\mu$ defined as follows:
\begin{definition}
	The cost regret $R^{\mu}$ of a given policy $\mu$ is defined as the difference between its expected total cost $C^{\mu}$ and the expected cost $C^{\mu_o}$ of the oracle policy $\mu_o$, i.e.,
	\begin{align}
		R^{\mu}=C^{\mu}-C^{\mu_o}.
	\end{align}
\end{definition}

In the next subsection, we study the cost regret $R^{\bar{\mu}}$ of our proposed low-complexity policy and the cost regret $R^{\mu^*}$ of the optimal policy along with other benchmarks such as two-step detect-then-intervene policies modified from QCD~\cite{veeravalli2014quickest}.

\subsection{Numerical Simulations}\label{sec:num_sim}
We compare the cost regret $R^\mu$ of the following four intervention policies, including our proposed low-complexity policy, optimal policy through grid approximation, and policies modified from quickest change detection algorithms \cite{veeravalli2014quickest}. These intervention policies are summarized as follows:
\begin{itemize}
	\item \textbf{Low-complexity:} The low-complexity policy is proposed in Algorithm \ref{algo:lowcomp}, which uses $\{\bar{\pi}_a\}$ as thresholds.
	\item \textbf{Optimal:} The optimal policy is obtained by grid approximation in Algorithm~\ref{algo:grid} with uniform and sufficiently small grids.
	\item \textbf{QCD:} The quickest change detection policy is a two-step intervention policy. It first detects the change point $\tau$ with Shiryaev's test \cite{veeravalli2014quickest} under an optimal chosen probability of false alarm (PFA). Then, it incrementally increases the intervention level to its maximum satisfying Eq.~\eqref{eq:consintervene}.
	\item \textbf{Direct QCD:} The direct quickest change detection policy is also a two-step policy similar to QCD. The only difference is that it violates constraint Eq.~\eqref{eq:consintervene} and immediately implements the highest intervention level $A$ once it claims the change-point has occurred.
\end{itemize}
Note that to verify the accuracy of approximations in the \emph{``local intervention''} regime, we also compare the regret for approximated cost $\tilde{C}$ computed from Corollary~\ref{cor:costbound}.

We simulated the diffusion process with five observation states $z_t\in\{1,2,3,4,5\}$, where larger $z_t$ represents a severer situation. The propagation costs associated with observations $z_t$ are $[0,1,2,3,4]$, where $z_t=1$ indicates a normal state where no propagation cost occurs. The actions which the agent can choose is $a_t\in\{0,1,2,3\}$, where $a_t=0$ represents idling and $a_t=3$ indicates perfect intervention. The intervention costs are $[0,0.02,0.06,0.2]$. As for the distribution of observations, we set the distribution before the change-point $\boldsymbol{\alpha}$ and that under the perfect intervention $\boldsymbol{\beta}_3$ to be $\boldsymbol{\alpha}=\boldsymbol{\beta}_3=[0.2,0.2,0.2,0.2,0.2]'$. For distributions under other intervention actions, we set $\boldsymbol{\beta}_i=[0.2-(6-2i)\delta,0.2-(3-i)\delta,0.3,0.2+(3-i)\delta,0.2+(6-2i)\delta]$, $\forall i \in\{0,1,2,3\}$. 

\begin{figure}
	\centering
	\subfigure[Low-complexity]{
		\centering
		\includegraphics[width=0.35\linewidth]{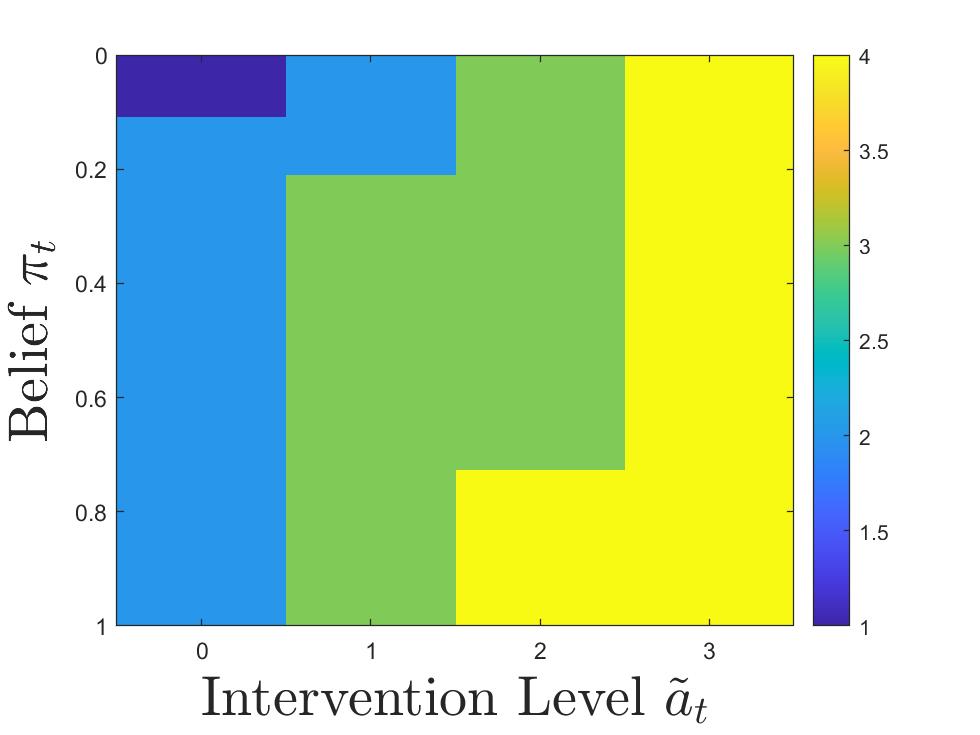}
	}
	\subfigure[Optimal]{
		\centering
		\includegraphics[width=0.35\linewidth]{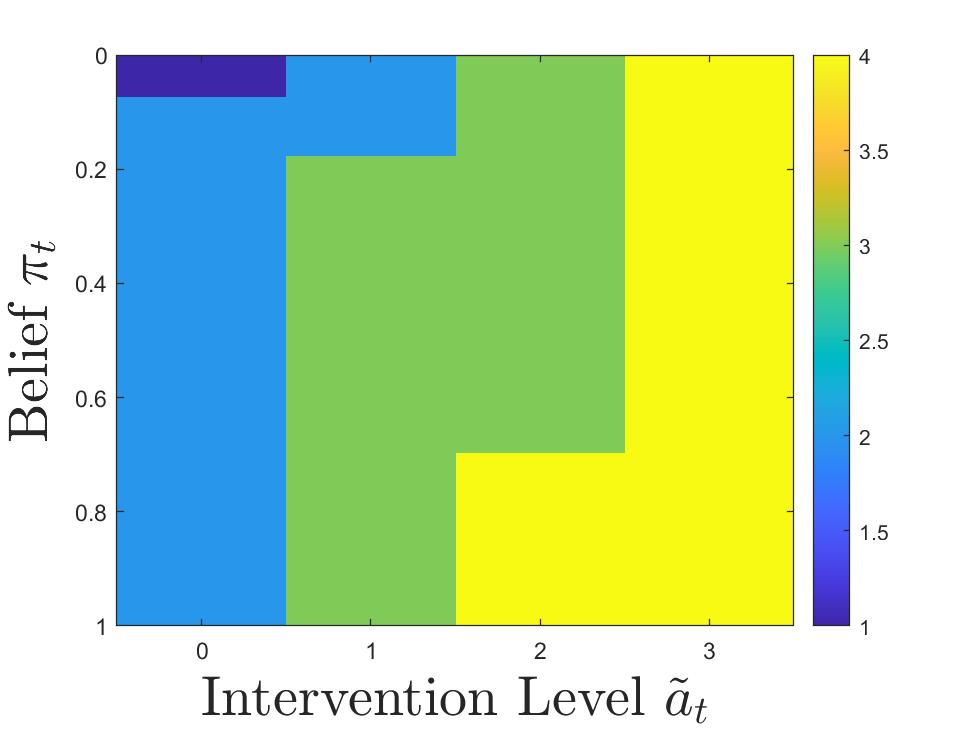}
	}
	\subfigure[QCD]{
		\centering
		\includegraphics[width=0.35\linewidth]{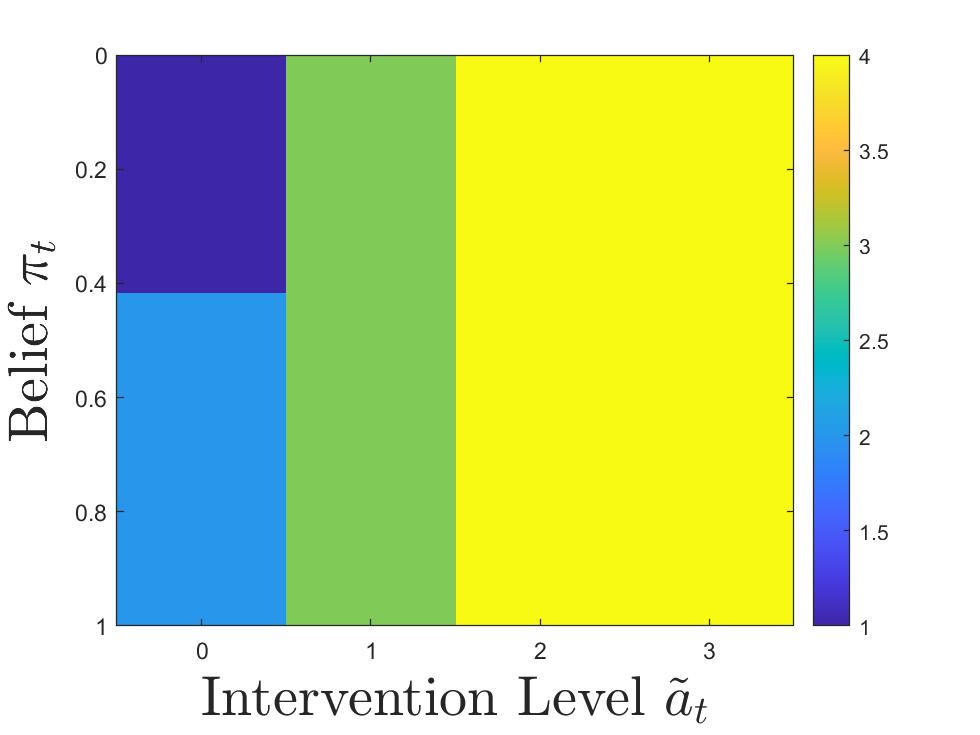}
	}
	\subfigure[DQCD]{
		\centering
		\includegraphics[width=0.35\linewidth]{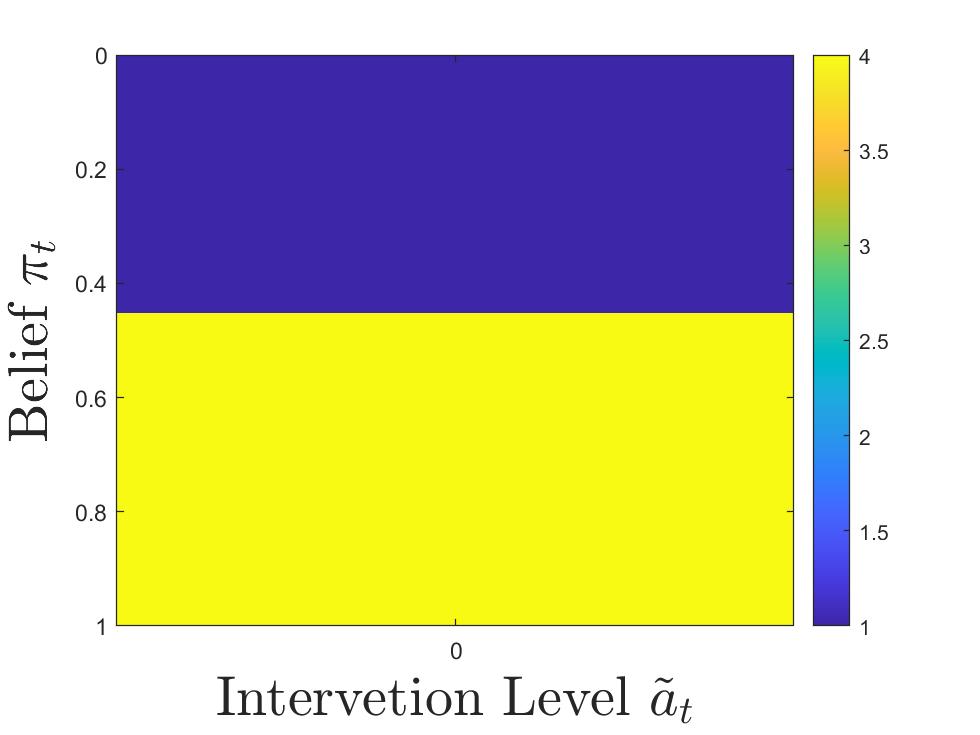}
	}
	\caption{Illustration of intervention policies: (a) proposed low-complexity policy, (b) optimal policy, (c) quickest change detection policy, and (d) direct quickest change detection policy. }
	\label{fig:policy}
\end{figure}

We first studied the threshold structure of these policies and the results are summarized  in Fig.~\ref{fig:policy}. Recall that $\rho$ is the geometric parameter that defines the distribution of horizon $T$, and $\lambda$ is the parameter that defines the geometric distribution of change-point $\tau$. Here, we set $\rho=0.99$ and $\lambda=0.03$. We further chose $\delta=0.02$. We see that the optimal policy possesses a threshold structure as we conjectured according to the results in the \emph{``local intervention''} regime. Low-complexity also has a threshold very close to Optimal.

\begin{figure}
	\centering
	\includegraphics[width=0.7\linewidth]{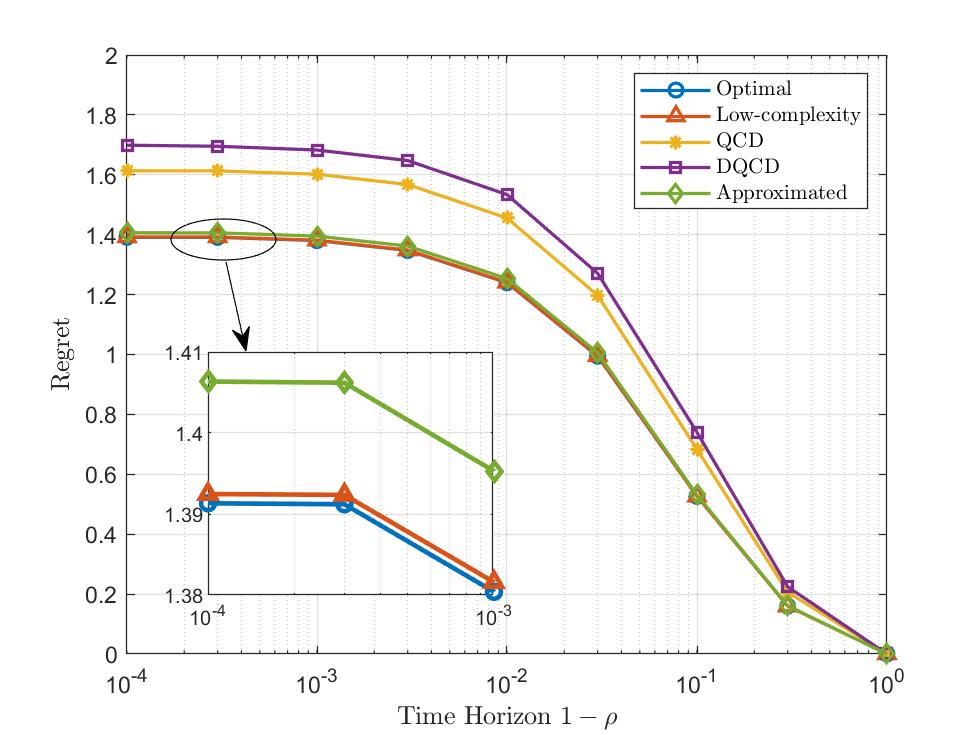}
	\caption{Cost regret performance as a function of time horizon parameter $1-\rho$.}
	\label{fig:Reg_rho}
\end{figure}

We then studied the performance of the four aforementioned policies under different parameter $\rho$, the geometric parameter that defines horizon $T$. Larger $1-\rho$ indicates that the diffusion process stops more quickly. Here, we set $\lambda=0.1$, $\delta=0.02$ and then let $1-\rho\in[10^{-4},1]$. The cost regret performance is depicted in Fig.~\ref{fig:Reg_rho}. It shows that as $\rho$ becomes smaller, the MDP stops with larger probability. Therefore, the probability that the agent needs to intervene also decreases. Low-complexity achieves nearly optimal regret performance under different $\rho$, and has much lower regret than the two QCD-based algorithms.
\begin{figure}
	\centering
	\includegraphics[width=0.7\linewidth]{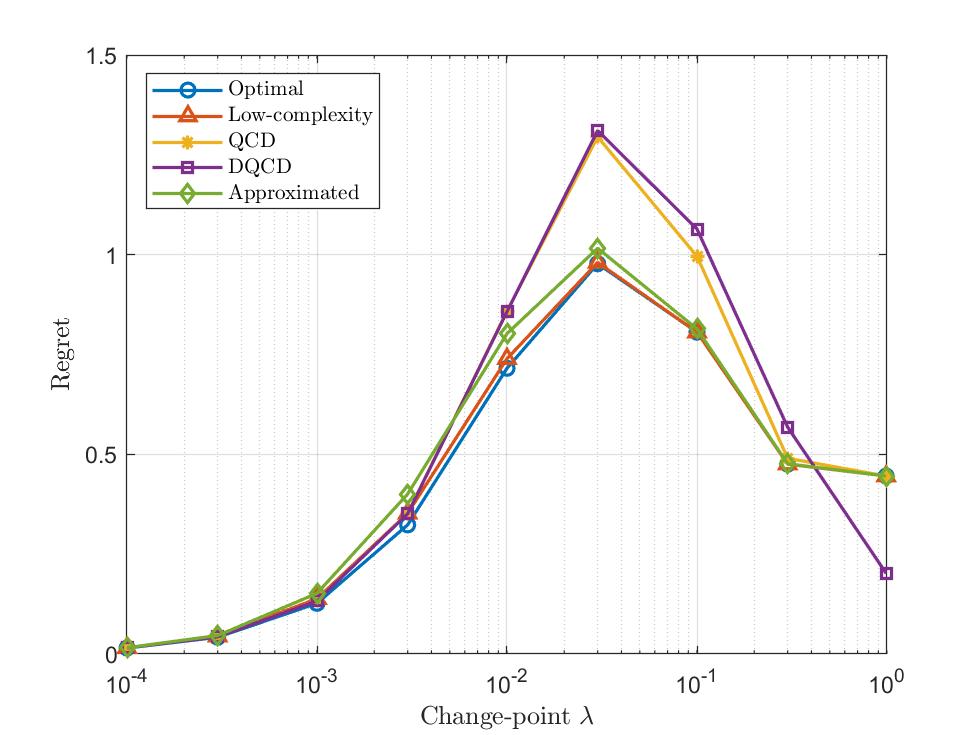}
	\caption{Cost regret performance as a function of change-point parameter $\lambda$. Note that DQCD performs even better than Optimal when $\lambda \rightarrow 1$ because it violates the constraint in Eq.~\eqref{eq:consintervene}. }
	\label{fig:Reg_lambda}
\end{figure}

Next, we studied the performance of four policies under different $\lambda$, the parameter that defines the geometric distribution of change-point. Here, we set $\rho=0.95$ and $\delta=0.2$, and then let $\lambda\in[10^{-4},1]$. From Fig.~\ref{fig:Reg_lambda}, we can see that our proposed low-complexity policy achieves nearly optimal performance when $\lambda$ is relatively large, but has a slightly larger regret than Optimal when $\lambda$ is small. This is because under such a regime, the upper bound $\bar{\pi}_a$ becomes loose. With $\lambda$ achieving $1$, the change-point is more likely to happen at time $1$. Since policy Direct QCD immediately switches to action $A$, violating the incremental increasing constraint in Eq.~\eqref{eq:consintervene}, it achieves the best performance even better than Optimal.
\begin{figure}
	\centering
	\includegraphics[width=0.7\linewidth]{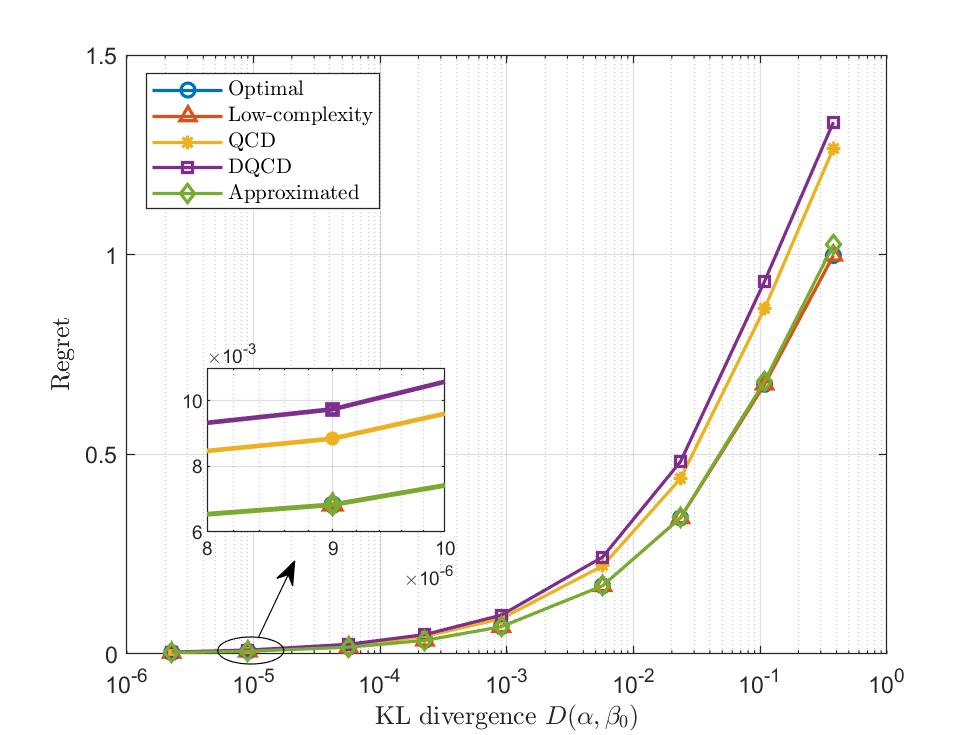}
	\caption{Cost regret performance as a function of KL divergence $D(\boldsymbol{\alpha},\boldsymbol{\beta}_0)$.}
	\label{fig:Reg_KL}
\end{figure}

Finally, we studied the cost performance of the four policies under different $\delta$. Since $\delta$ is viewed as the distance between consecutive intervention action distributions, it can be replaced by the KL divergence of $\boldsymbol{\alpha}$ and $\boldsymbol{\beta}_0$ for better understanding. Here, we set $\rho=0.95$, $\lambda=0.1$, and let $\delta\in[10^{-4},0.03]$. The KL divergence $D(\alpha,\beta_0)$ lies within $[10^{-6},1]$. It shows that Low-complexity achieves nearly optimal performance under different $\delta$. As a matter of fact, when the distributions are closer to each other, i.e., in the \emph{``local intervention''} regime, the improvement of our low-complexity policy compared to QCD based policies becomes larger. The regret decreases by around $22\%$ when the KL divergence is as small as $\mathcal{O}(10^{-6})$.

Throughout the simulations, our approximated cost value $\tilde{C}$ is very close to the cost $C^{\mu^*}$ of the optimal policy. From Fig.~\ref{fig:Reg_KL}, we see that the approximated cost regret only deviates from the regret of Optimal when the KL divergence $D(\alpha,\beta_0)$ deviates significantly from the \emph{``local intervention''} regime. Otherwise, the approximation is very accurate.

\subsection{Application to Anomaly Detection}
\begin{figure}
    \centering
    \includegraphics[width=0.7\linewidth]{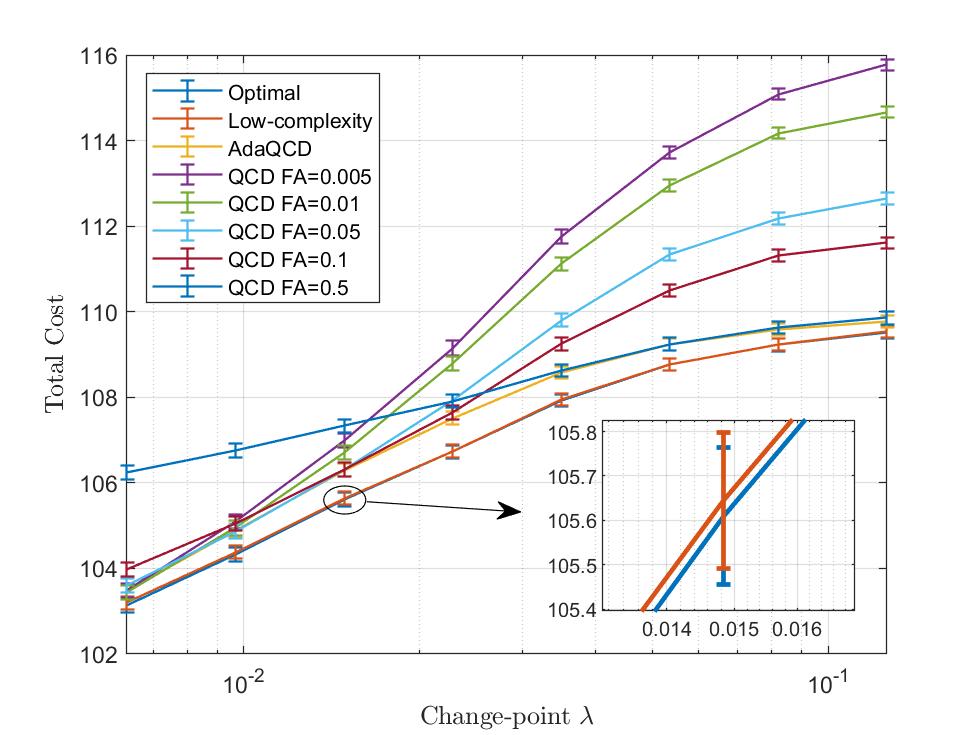}
    \caption{Total cost performance as a function of change-point parameter $\lambda$.}
    \label{fig:exp_lambda}
\end{figure}

In this subsection, we apply our low-complexity policy in algorithm~\ref{algo:lowcomp} to a simple anomaly detection problem. Consider a scenario where a communication link is provided by a service provider to users for packets transmission. For simplicity, we assume at each time $t$, a normal user chooses among five different service rates (denoted as $z_t\in\{1,2,3,4,5\}$ from low speed to high speed) to transmit these packets, equally likely (this assumption can be easily generalized). The user may become malicious after some change-points $\tau$ and start to send with an unusually high rate for a long period of time. The service provider monitors the user's transmission rate and charge the user accordingly. Based on the observations, the provider needs to decide whether the user has become malicious. If so, it may take intervention actions to throttle the data rate. The scenario described above can be viewed as a simplification of general MAC layer anomaly detection~\cite{CardenasMAC}. We assume that the service costs associated with different speeds, from low to high, are $\{0,1,2,3,4\}$ and the provider has four actions ($a_t \in\{0,1,2,3\}$ where $0$ represents idle) to choose from, which increasingly limit the high-speed transmission rate with intervention costs $\{0, 0.02, 0.06, 0.2\}$. The probability distributions of the service rate under different intervention actions are the same as in Sec.~\ref{sec:num_sim} with $\delta = 0.02$. However, we assume a fixed time horizon $T = 50$ in our experiment, and the service provider can choose any action at any time with no restriction such as Eq.~\eqref{eq:consintervene}. In order to compute the low-complexity policy, we set $\rho = 0.98$. We simulated this system $20,000$ times and compared the performance of our low-complexity policy with the following baselines: (1) optimal policy computed through dynamic programming; (2) QCD policy with fixed PFA; (3) QCD policy with adaptively chosen optimal PFA. The total cost performance under different change-point parameter $\lambda$ is shown in Fig.~\ref{fig:exp_lambda}. It turns out that the total cost performance of our proposed low-complexity policy is nearly optimal, in the sense that its mean is close to the optimal policy and their confidence intervals are intersecting one another. A significant performance gap can be observed between the best QCD policy and the low-complexity policy, which demonstrates the superiority of our proposed policy. This result also shows that leveraging the assumptions such as Eq.~\eqref{eq:consintervene} in our technical sections will not harm the total cost performance of our proposed low-complexity policy too much, even though it is designed based on such assumptions.

\section{Conclusion}\label{sec:conclusion}
In this work, we studied the problem of controlling a diffusion process with mutation. We formulated the problem as a partially observed MDP (POMDP) and converted it into a fully observed MDP with the formulation of belief states. We first proposed a grid approximation algorithm to compute the optimal intervention policy. To reduce computational complexity, we considered the \emph{``local intervention''} regime and approximated the value functions through first-order Taylor's expansion under this regime. We then proposed a low-complexity policy based on the upper bounds of belief thresholds derived from the approximation. Finally, Simulation and experiment results have been presented to verify the performance of the proposed intervention policies, which showed that the low-complexity algorithm has a similar regret as the optimal policy even beyond the \emph{``local intervention''} regime. 

\bibliographystyle{IEEEtran}
\bibliography{ref, inlab-refs}


\clearpage
\appendix

\section*{Proof of Theorem \ref{theo:beliefMDP}}
Consider the cost objective in Problem~\ref{pb:costmin} and apply the law of total expectation conditioned on the time horizon $T$ as follows:
\begin{align}\label{eq:beliefMDP1}
	C^\mu=&\mathbb{E}\left[c_i^{a_0}+\sum_{t=1}^T(c_p^{z_t}+c_i^{a_t})\right]\nonumber\\
	=& \mathbb{E}_T\left[\left.\mathbb{E}\left[c_i^{a_0}\right]+\sum_{t=1}^K\mathbb{E}\left[c_p^{z_t}+c_i^{a_t}\right]\right|T=K\right].
\end{align}
For a fixed time $t$, we apply again the law of total expectation conditioned on the filtration $\mathcal{F}_t$ as follows:
\begin{align*}
    \mathbb{E}\left[c_p^{z_t}+c_i^{a_t}\right]
	=&\mathbb{E}_{\mathcal{F}_t}\left[\mathbb{E}\left[c_p^{z_t}+c_i^{a_t}|\mathcal{F}_t\right]\right].
\end{align*}
Through minimizing Eq.~\eqref{eq:beliefMDP1}, we will obtain a function $a_t=a(\mathcal{F}_t)$ as solution. However, since according to Theorem 7.2.1 in \cite{krishnamurthy2016partially}, $\pi_t$ is a sufficient statistic for $\mathcal{F}_t$, therefore we have:
\begin{align}
	\mathbb{E}[c_p^{z_t}+c_i^{a_t}|\mathcal{F}_t]=&c_p^{z_t}+\sum_{a}\mathrm{Pr}(a_t=a|\mathcal{F}_t)c_i^a\nonumber\\
	=&c_p^{z_t}+\sum_{a}\mathrm{Pr}(a_t=a|\pi_t)c_i^a\nonumber\\
	=& \mathbb{E}[c_p^{z_t}+c_i^{a_t}|\pi_t].
\end{align}
So the optimal action is given by $a_t=a(\pi_t)$. Therefore, the POMDP is converted into a fully observed MDP with state $(\pi_t,\tilde{a}_t,z_t)$.

\section*{Proof of Lemma \ref{lemma:concavity}}
In the proof, we use superscript $^{(k)}$ to denote the $k$-th iteration of standard value iteration process. Then, according to Bellman equation~\eqref{eq:bellman}, standard value iteration process can be written as:
\begin{subequations}\label{eq:concavitybellman}
\begin{align}
	J_a^{(k+1)}(\pi)=&c_i^a+\rho\sum_{z=1}^Z\sigma_a(\pi,z)\left[c_p^z+V_a^{(k)}(T_a(\pi,z))\right],\\
	V_{\tilde{a}}^{(k+1)}(\pi)=&\min\{J_{\tilde{a}}^{(k)}(\pi),J_{\min\{\tilde{a}+1,A\}}^{(k)}(\pi)\}.
\end{align}
\end{subequations}

We first prove the concavity by induction. By the contraction property of standard value iteration process, we can choose $V_{\tilde{a}}^{(0)}(\pi)=J_a^{(0)}(\pi)=0$ as base case. Then, they are both concave.

Suppose at iteration $k$, the value functions $V_{\tilde{a}}^{(k)}(\pi)$ and $J_a^{(k)}(\pi)$ are all concave. Consider $\pi_1$ and $\pi_2$ are two different values between $[0,1]$. Without loss of generality, suppose $\pi_1<\pi_2$. Define $\pi_3=\lambda\pi_1+(1-\lambda)\pi_2$, where $\lambda\in[0,1]$. For simplicity, we denote
\begin{align}\label{eq:concavity1}
	\sigma=\frac{\lambda\sigma_a(\pi_1,z)}{\lambda\sigma_a(\pi_1,z)+(1-\lambda)\sigma_a(\pi_2,z)}.
\end{align}
It is easy to verify that:
\begin{align}\label{eq:concavity2}
	T_a(\pi_3,z)=\sigma T_a(\pi_1,z) +(1-\sigma)T_a(\pi_2,z).
\end{align}

Notice that according to Eq.~\eqref{eq:concavitybellman}, for fixed $a$ and iteration $k$, $J_a^{(k+1)}(\pi)$ can be divided into three terms as follows:
\begin{align}
    J_a^{(k+1)}(\pi)=c_i^a+\rho \hat{C}_p(\pi)+\rho\hat{V}(\pi),
\end{align}
where 
\begin{align*}
    \hat{C}_p(\pi)=&\sum_{z=1}^Z\sigma_a(\pi,z)c_p^z,\\
    \hat{V}(\pi)=&\sum_{z=1}^Z\sigma_a(\pi,z)V_a^{(k)}(T_a(\pi,z)).
\end{align*}

Therefore, the analysis of linear combination of $J_a^{(k+1)}(\pi_1)$ and $J_a^{(k+1)}(\pi_2)$ with parameter $\lambda$ can also be divided into two parts. Since $c_i^a+\rho\hat{C}_p(\pi)$ is affine, we will first have:
\begin{align}\label{eq:concavity3}
    \lambda J_a^{(k+1)}(\pi_1)+(1-\lambda)J_a^{(k+1)}(\pi_2)
    =c_i^a+\rho\hat{C}_p(\pi_3)+\rho\left[\lambda\hat{V}(\pi_1)+(1-\lambda)\hat{V}(\pi_2)\right].
\end{align}

Then, we analyze the linear combination of $\hat{V}(\pi)$ as follows. Note that for simplicity, we omit $z$ and $a$ when there is no obscure.
\begin{align}
    \lambda\hat{V}(\pi_1)+(1-\lambda)\hat{V}(\pi_2)
    =&\sum_{z'=1}^Z\left[\lambda\sigma_a(\pi_1)V^{(k)}(T(\pi_1))+(1-\lambda)\sigma_a(\pi_2)V^{(k)}(T(\pi_2))\right]\nonumber\\
    \overset{(a)}{=}& \sum_{z'=1}^Z\sigma_a(\pi_3,z)\left[\sigma V^{(k)}(T(\pi_1))+(1-\sigma)V^{(k)}(T(\pi_2))\right]\nonumber\\
    \overset{(b)}{\leq}&\sum_{z'=1}^Z\sigma_a(\pi_3,z)V^{(k)}(\sigma T(\pi_1)+(1-\sigma)T(\pi_2))\nonumber\\
    \overset{(c)}{=}&\sum_{z'=1}^Z\sigma_a(\pi_3,z)V^{(k)}(T(\pi_3))\nonumber\\
    =&\hat{V}(\pi_3),\label{eq:concavity4}
\end{align}
where equality $(a)$ is due to the definition of $\sigma$ in Eq.~\eqref{eq:concavity1} and the fact that $\sigma_a(\pi,z)$ is linear with $\pi$. Inequality $(b)$ is due to the concavity assumption of $V^{(k)}_a(\pi)$, and equality $(c)$ is due to Eq.~\eqref{eq:concavity2}. 

By substituting Eq.~\eqref{eq:concavity3} into Eq.~\eqref{eq:concavity4}, we have:
\begin{align*}
    \lambda J_a^{(k+1)}(\pi_1)+(1-\lambda)J_a^{(k+1)}(\pi_2)
	\leq & c_i^a+\rho \hat{C}_p(\pi_3)+\rho \hat{V}(\pi_3)\nonumber\\
	=& J_a^{(k+1)}(\pi_3).
\end{align*}

So the concavity of $J_a^{(k+1)}(\pi)$ is proved. Since $V_{\tilde{a}}^{(k+1)}(\pi)$ is the minimum of two concave functions according to Eq.~\eqref{eq:concavitybellman}, it is also concave. Let $k\rightarrow\infty$, we can prove the concavity of $J_a(\pi)$ and $V_{\tilde{a}}(\pi)$.

Specifically according to the Bellman equation \eqref{eq:concavitybellman}, when considering $a=A$, we have:
\begin{align}
	J_A^{(k+1)}(\pi)=c_i^a+\rho\sum_{z=1}^Z\alpha(z)c_p^z+\rho J_A^{(k)}(\tilde{\pi}).
\end{align}

From a similar induction argument, we can prove that $J_A(\pi)$ is constant. Suppose $J_A^{(k)}(\tilde{\pi})$ is a constant, then $J_A^{(k+1)}(\pi)$ is obviously a constant. Therefore, $J_A(\pi)$ is a constant. Set $J_A^{(k)}(\tilde{\pi})=J_A^{(k+1)}(\pi)$ and we get:
\begin{align}
	J_A(\pi)=\frac{c_i^a+\rho\sum_{z=1}^Z\alpha(z)c_p^z}{1-\rho}.
\end{align}

Next, we prove the monotonic property. First, we can easily verify that $T_a(\pi,z)$ is monotonically non-decreasing with $\pi$. Then, when $\boldsymbol{\beta}_a$ and $\boldsymbol{\alpha}$ satisfies MLR ordering $\boldsymbol{\beta}_a\geq_r\boldsymbol{\alpha}$, we can also easily check that $T_a(\pi,z)$ is monotonically non-decreasing with $z$. Suppose $\overline{\pi}>\pi$, we can easily check:
\begin{align*}
	\sum_{z=j}^Z\sigma_a(\overline{\pi},z)\geq \sum_{z=j}^Z\sigma_a(\pi,z), \quad \forall j\in\{1,2,\cdots,Z\},
\end{align*}
which means stochastic dominance $\boldsymbol{\sigma}_a(\overline{\pi})\geq_s\boldsymbol{\sigma}_a(\pi)$. According to its properties \cite{krishnamurthy2016partially}, if $V_a(T_a(\pi,z))$ is monotonically non-decreasing with $z$, we have:
\begin{align*}
	\sum_{z=1}^Z\sigma_a(\overline{\pi},z)V_a(T_a(\pi,z))\geq\sum_{z=1}^Z\sigma_a(\pi,z)V_a(T_a(\pi,z)).
\end{align*}

Based on a similar induction argument, if $V_a^{(k)}(\pi)$ is monotonically non-decreasing with $\pi$, we have $V_a^{(k)}(T_a(\pi,z))$ is monotonically non-decreasing with $\pi$ since $T_a(\pi,z)$ is non-decreasing with $\pi$. Also notice that $V_a^{(k)}(T_a(\pi,z))$ is monotonically non-decreasing with $z$ since $T_a(\pi,z)$ is non-decreasing with $z$. Then, by the property of stochastic dominance, we will get:
\begin{align*}
	\sum_{z=1}^Z\sigma_a(\overline{\pi},z)V_a^{(k)}(T_a(\overline{\pi},z))\geq\sum_{z=1}^Z\sigma_a(\pi,z)V_a^{(k)}(T_a(\pi,z)).
\end{align*}

With a little manipulation, we can get $J_a^{(k+1)}(\overline{\pi})\geq J_a^{(k+1)}(\pi)$, which means $J_a^{(k+1)}(\pi)$ is non-decreasing. Since $V_{\tilde{a}}^{(k+1)}(\pi)$ is the minimum of non-decreasing functions, it is also monotonically non-decreasing. Let $k\rightarrow\infty$, we prove the monotonicity of $J_a(\pi)$ and $V_{\tilde{a}}(\pi)$. 

\section*{Proof of Theorem \ref{theo:thressmall}}
To prove this theorem, we first need a lemma that states the concavity and monotonicity the approximated value function similar to Lemma~\ref{lemma:concavity}. The proof of this lemma follows exactly the same induction procedure as in the proof of Lemma~\ref{lemma:concavity}, thus it is omitted.
\begin{lemma}\label{lemma:concavityapprox}
    For any action $a$ and current intervention level $\tilde{a}$, the approximated value function $\tilde{V}_a(\pi)$ and the action-value function $\tilde{J}_a(\pi)$ are both concave and monotonically non-decreasing in belief state $\pi$. Furthermore, $\tilde{V}_A(\pi)$ and $\tilde{J}_A(\pi)$ are constant: 
	\begin{align}
		\tilde{V}_A(\pi)=\tilde{J}_A(\pi)=\frac{c_i^A+\rho\sum_{z=1}^Z\alpha(z)c_p^z}{1-\rho}.
	\end{align}
\end{lemma}

With Lemma~\ref{lemma:concavityapprox}, we prove Theorem~\ref{theo:thressmall} through induction following standard value iteration. We first prove the monotonicity of $\tilde{\Delta}_a^{\delta}(\pi)$. In the rest of the proof, we omit superscript $\delta$ without obscurity and put value iteration superscript $^{(k)}$ into the equation. Recall that from Eq.~\eqref{eq:bellmanapproxdiff}:
\begin{align*}
    \tilde{\Delta}_a^{(k+1)}(\pi)=&\nonumber
	\tilde{J}_{a+1}^{(k+1)}(\pi)-\tilde{J}_a^{(k+1)}(\pi)\\
	=&D_i^{a+1}+\tilde{\pi}\rho D_p^{a+1}+\rho\left[\tilde{V}_{a+1}^{(k)}(\tilde{\pi})-\tilde{V}_{a}^{(k)}(\tilde{\pi})\right].\nonumber
\end{align*}
Notice that $$D_p^{a+1}=\sum_{z=1}^Z(\beta_{a+1}(z)-\beta_a(z))c_p^z<0$$ since $\boldsymbol{\beta_a}\geq_s\boldsymbol{\beta_a}$ as the intervention level assumption. So $D_i^{a+1}+\tilde{\pi}\rho D_p^{a+1}$ is non-increasing with $\pi$.

According to the Bellman equation, $$\tilde{V}_{a+1}^{(k)}=\min\{\tilde{J}_a^{(k)}(\pi),\tilde{J}_{\min\{a+1,A\}}^{(k)}(\pi)\}.$$

If $a=A-1$, we have:
\begin{align*}
	\tilde{V}_{a+1}^{(k)}-\tilde{V}_{a}^{(k)}=&\tilde{J}_A^{(k)}(\pi)-\min\{\tilde{J}_{A-1}^{(k)}(\pi),\tilde{J}_{A}^{(k)}(\pi)\}\nonumber\\
	=&\max\{\tilde{J}_{A}^{(k)}(\pi)-\tilde{J}_{A-1}^{(k)}(\pi),0\}\nonumber\\
	=&\max\{\Delta^{(k)}_{A-1}(\pi),0\}.
\end{align*}

By induction assumption, $\Delta^{(k)}_{A-1}(\pi)$ is non-increasing. Then, we prove that $\tilde{V}_{A}^{(k)}-\tilde{V}_{A-1}^{(k)}$ is non-increasing. As a result, $\tilde{\Delta}_{A-1}^{(k+1)}(\pi)$ is non-increasing. 

If $a\neq A-1$, we have:
\begin{align*}
	\tilde{V}_{a+1}^{(k)}-\tilde{V}_{a}^{(k)}=&\min\left\{\tilde{J}_{a+2}^{(k)}(\pi),\tilde{J}_{a+1}^{(k)}(\pi)\right\}-\min\left\{\tilde{J}_{a+1}^{(k)}(\pi),\tilde{J}_{a}^{(k)}(\pi)\right\}.
\end{align*}

The result could be one of the following four terms: $0$, $\Delta_{a+1}^{(k)}(\pi)$, $\Delta_{a}^{(k)}(\pi)$, or $\Delta_{a+1}^{(k)}(\pi)+\Delta_{a}^{(k)}(\pi)$. According to the induction assumption, these three terms are all monotonically non-increasing, which means that $\tilde{V}_{a+1}^{(k)}-\tilde{V}_{a}^{(k)}$ is monotonically non-increasing for any $a$. Therefore, $\tilde{\Delta}^{(k+1)}_a(\pi)$ is monotonically non-increasing with $\pi$ for any $a$. Let $k\rightarrow\infty$, we prove the monotonicity of $\tilde{\Delta}_a(\pi)$.

According to the definition of submodular functions and nearly-submodular functions in Definition~\ref{def:submodular} and Definition~\ref{def:nearlysubmodular} respectively, we prove that the original action-value function $J_a^{\delta}(\pi)$ is nearly-submodular in pair $(a,\pi)$, and the approximated action-value function $\tilde{J}_a^{\delta}(\pi)$ is submodular.

Denote $\tilde{\mu}^{*,\delta}(\pi)$ to be the optimal policy that solves the approximated Bellman equations \eqref{eq:bellmanapprox}. Since $\tilde{J}_a(\pi)$ is submodular, according to Topkis' lemma \cite{Topkis78submodular}, $\tilde{\mu}^{*,\delta}(\pi)$ is monotonically non-decreasing. Due to the finiteness of action space of $a$, there exists a threshold structure $\{\tilde{\pi}_a^{*,\delta}\}_{1\leq a\leq A}$, such that policy $\tilde{\mu}^{*,\delta}(\pi)$ switches from action $a$ to a stricter action $a+1$ once $\pi$ exceeds threshold $\tilde{\pi}_a^{*,\delta}$.

\section*{Proof of Corollary \ref{cor:costbound}}
In the proof, we omit superscript $\delta$ when there is no obscurity. According to the Bellman equation for approximated value function Eq.~\eqref{eq:bellmanapprox}, we have:
\begin{align*}
	\tilde{J}_a(\pi)= c_i^a+\rho \sum_{z=1}^Z \sigma_a(\pi,z)c_p^z+\rho \tilde{V}_a(\tilde{\pi}).
\end{align*}

From this approximated Bellman equation, suppose at step $t$, the belief is $\pi_t$. At the next step, the belief will evolve to $\pi_{t+1}=\pi_t+\lambda(1-\pi_t)$. The dynamics of belief update provides us with the general term formula as:
\begin{align}
	\pi_t=(1-\lambda)^{t}(\pi_0-1)+1.
\end{align}
If the initial belief $\pi_0=0$, the formula will reduce to $\pi_t=1-(1-\lambda)^{t}$.

According to Theorem~\ref{theo:thressmall}, the optimal policy $\tilde{\mu}^*(\pi)$ possesses a threshold structure $\{\tilde{\pi}_a^*\}_{1\leq a\leq A}$. Therefore, between thresholds $\tilde{\pi}_a^*$ and $\tilde{\pi}_{a+1}^*$ we always choose the same action $a$. At this stage, the expected cost $\tilde{V}_a(\pi_0)$ with initial belief $\pi_0\in[\tilde{\pi}_a^*,\tilde{\pi}_{a+1}^*)$ can be bounded as:
\begin{align*}
	\tilde{V}_a(\pi_0)=& c_i^a+\rho \sum_{z=1}^Z\alpha(z)c_p^z+\rho\pi_1A_p^a+\rho \tilde{V}_a(\pi_1)\nonumber\\
	= & \sum_{\tau=1}^{t}\left[\rho^{\tau-1}c_i^a+\rho^{\tau} \left(B_p+\pi_{\tau}A_p^a\right)\right]+\rho^t\tilde{V}_a(\pi_t),
\end{align*}
where $B_p=\sum_{z=1}^Z\alpha(z)c_p^z$ and if belief $\pi_{t-1}\in[\tilde{\pi}_a^*,\tilde{\pi}_{a+1}^*)$.

Define $t_a$ be the first time that $\pi_t\geq \tilde{\pi}_a^*$. Then, we have:
\begin{align*}
    t_a=\left\lceil\frac{\ln(1-\tilde{\pi}_a^*)}{\ln(1-\lambda)}\right\rceil.
\end{align*}

Before time $t_1$, the optimal action is to idle with $a=0$. Then, we can bound the total approximated cost $\tilde{C}$ as:
\begin{align}
    \tilde{C}=&\tilde{V}_0(0)\nonumber\\
    = & \sum_{\tau=1}^{t_1}\left[\rho^{\tau-1}c_i^0+\rho^{\tau}\left(B_p+\pi_{\tau}A_p^0\right)\right]+\rho^{t_1} \tilde{V}_0(\pi_{t_1})\nonumber\\
    =& \sum_{a=0}^{A-1}\sum_{\tau=t_a+1}^{t_{a+1}}\left[\rho^{\tau-1}c_i^a+\rho^{\tau}\left(B_p+\pi_{\tau}A_p^a\right)\right]+\rho^{t_A}V_A\nonumber\\
    =&\underbrace{\sum_{a=0}^{A-1}\sum_{\tau=t_a+1}^{t_{a+1}}\rho^{\tau-1}c_i^a}_{F_1} +\underbrace{\sum_{a=0}^{A-1}\sum_{\tau=t_a+1}^{t_{a+1}}\rho^{\tau}B_p}_{F_2}+ \underbrace{\sum_{a=0}^{A-1}\sum_{\tau=t_a+1}^{t_{a+1}}\rho^{\tau}\pi_{\tau}A_p^a}_{F_3}
    +\rho^{t_A}V_A,
\end{align}
where $t_0=0$ and $V_A=\frac{c_i^A+\rho\sum_{z=1}^Z\alpha(z)c_p^z}{1-\rho}$ is a constant according to Lemma~\ref{lemma:concavityapprox}. First, we calculate the first term $F_1$ as follows:
\begin{align*}
    F_1=&\sum_{a=0}^{A-1}\frac{1}{1-\rho}\left(\rho^{t_a}c_i^a-\rho^{t_{a+1}}c_i^a\right)\nonumber=\frac{1}{1-\rho}\left(\sum_{a=0}^{A-1}\rho^{t_a}c_i^{a}-\sum_{a=1}^{A}\rho^{t_a}c_i^{a-1}\right)\nonumber\\
    =& \frac{1}{1-\rho}\left(\sum_{a=1}^A\rho^{t_a}D_i^a-\rho^{t_A}c_i^A\right).
\end{align*}
Then, we calculate the second term $F_2$ as follows:
\begin{align*}
    F_2=&\sum_{a=0}^{A-1}\sum_{\tau=t_a+1}^{t_{a+1}}\rho^{\tau}B_p=\frac{\rho}{1-\rho}\left(1-\rho^{t_A}\right)B_p.
\end{align*}
The third term $F_3$ can also be calculated as follows:
\begin{align*}
    F_3=&\sum_{a=0}^{A-1}\sum_{\tau=t_a+1}^{t_{a+1}}\rho^{\tau}\pi_{\tau}A_p^a\nonumber
    =\sum_{a=0}^{A-1}\sum_{\tau=t_a+1}^{t_{a+1}}\rho^{\tau}(1-(1-\lambda)^\tau)A_p^a\nonumber\\
    =&\underbrace{\sum_{a=0}^{A-1}\sum_{\tau=t_a+1}^{t_{a+1}}\rho^{\tau}A_p^a}_{F_{31}}-\underbrace{\sum_{a=0}^{A-1}\sum_{\tau=t_a+1}^{t_{a+1}}[\rho(1-\lambda)]^\tau A_p^a}_{F_{32}}.
\end{align*}
The first term $F_{31}$ in the above equation can be simplified as:
\begin{align*}
    F_{31}=&\sum_{a=0}^{A-1}\sum_{\tau=t_a+1}^{t_{a+1}}\rho^{\tau}A_p^a=\frac{\rho}{1-\rho}\sum_{a=0}^{A-1}\left(\rho^{t_a}A_p^a-\rho^{t_{a+1}}A_p^a\right)\nonumber
    =\frac{\rho}{1-\rho}\left(A_p^0+\sum_{a=1}^A\rho^{t_a}D_p^a\right),
\end{align*}
and the second term $F_{32}$ is simplified as:
\begin{align*}
    F_{32}=\sum_{a=0}^{A-1}&\sum_{\tau=t_a+1}^{t_{a+1}}[\rho(1-\lambda)]^\tau A_p^a
    =\frac{\rho(1-\lambda)}{1-\rho(1-\lambda)}\left(A_p^0+\sum_{a=1}^A[\rho(1-\lambda)]^{t_a}D_p^a\right).
\end{align*}
Therefore, $F_3$ is simplified as:
\begin{align*}
    F_3=F_{31}+F_{32}=\frac{\lambda\rho }{(1-\rho)[1-\rho(1-\lambda)]}A_p^0+\sum_{a=1}^A\left(\frac{\rho^{t_a+1}}{1-\rho}-\frac{[\rho(1-\lambda)]^{t_a+1}}{1-\rho(1-\lambda)}\right)D_p^a.
\end{align*}

Combining $F_1$, $F_2$ and $F_3$ with $V_A$, we have the bound on $\tilde{C}$ as:
\begin{align}
    \tilde{C}=&F_1+F_2+F_3+\rho^{t_A}V_A\nonumber\\
    =&\sum_{a=1}^A\frac{\rho^{t_a}D_i^a}{1-\rho}+\sum_{a=1}^A\left(\frac{\rho^{t_a+1}}{1-\rho}-\frac{[\rho(1-\lambda)]^{t_a+1}}{1-\rho(1-\lambda)}\right)D_p^a +\frac{\rho\sum_{z=1}^Z\alpha(z)c_p^z}{1-\rho}+\frac{\lambda\rho A_p^0}{(1-\rho)[1-\rho(1-\lambda)]}.
\end{align}

\section*{Proof of Corollary \ref{cor:thresboundsmall}}
The threshold $\tilde{\pi}_a^{*,\delta}$ is obtained through setting $\tilde{\Delta}_{a}^\delta(\pi)=0$. In the proof, we omit superscript $^\delta$ for simplicity. We have:
\begin{align*}
	0=\tilde{\Delta}_{a}^\delta(\pi)=D_i^a+\tilde{\pi}\rho D_p^a+\rho\left[\tilde{V}_{a+1}(\tilde{\pi})-\tilde{V}_{a}(\tilde{\pi})\right].\nonumber
\end{align*}

Through simple manipulation, we can get the expression of $\tilde{\pi}$ as:
\begin{align*}
	\tilde{\pi}=\frac{D_i^a}{-\rho D_p^a}+\frac{\tilde{V}_{a+1}(\tilde{\pi})-\tilde{V}_{a}(\tilde{\pi})}{-D_p^a}.
\end{align*}

Recall that $\tilde{\Delta}_{a}^\delta(\pi)$ is monotonically non-increasing according to Theorem~\ref{theo:thressmall}. When setting $\pi=\tilde{\pi}_a^*$, the sudo-belief $\tilde{\pi}=\pi+\lambda(1-\pi)$ will exceed the threshold $\tilde{\pi}_a^*$. Therefore, we have $\tilde{\Delta}_{a}(\tilde{\pi}) \leq 0$ and $V_a(\tilde{\pi})=J_{a+1}(\tilde{\pi})$. So we have:
\begin{align*}
	\tilde{\pi}=&\frac{D_i^a}{-\rho D_p^a}+\frac{\min\{\tilde{J}_{a+1}(\tilde{\pi}),\tilde{J}_{a+2}(\tilde{\pi})\}-\tilde{J}_{a+1}(\tilde{\pi})}{-D_p^a}\nonumber\\
	=&\frac{D_i^a}{-\rho D_p^a}+\frac{\min\{0,\tilde{\Delta}_{a+1}(\tilde{\pi})\}}{-D_p^a}.
\end{align*}
Therefore, we have:
\begin{align}
	\tilde{\pi}\leq-\frac{D_i^a}{\rho D_p^a}.
\end{align}

The upper bound $\tilde{\pi}_a^{*,\delta}\leq \bar{\pi}$ is obtained simply from substituting $\tilde{\pi}=\tilde{\pi}_a^{*,\delta}+\lambda(1-\tilde{\pi}_a^{*,\delta})$ into the above equation.

Next, we consider the lower bound. Since $\tilde{\Delta}_{a+1}(\tilde{\pi})$ is non-increasing, we have $\tilde{\Delta}_{a+1}(\tilde{\pi})\geq\tilde{\Delta}_{a+1}(1)$. Since when $\pi=1$, which means the change-point has already happened almost surely, taking a stricter action $a+1$ will be always better. Then, we have $\Delta_{a+1}(1)<0$. So, we have:
\begin{align}\label{eq:thresboundsmall1}
	\tilde{\pi}=&\frac{D_i^a}{-\rho D_p^a}+\frac{\min\{0,\tilde{\Delta}_{a+1}(\tilde{\pi})\}}{-D_p^a}\nonumber\\
	\geq&\frac{D_i^a}{-\rho D_p^a}+\frac{\tilde{\Delta}_{a+1}(1)}{-D_p^a}.
\end{align}

Recall that:
\begin{align*}
	\tilde{\Delta}_{a+1}(1)=&D_i^{a+1}+\rho D_p^{a+1}+\rho\left[\tilde{V}_{a+2}(1)-\tilde{V}_{a+1}(1)\right]\nonumber\\
	=& D_i^{a+1}+\rho D_p^{a+1}+\rho\left[\tilde{J}_{a+3}(1)-\tilde{J}_{a+2}(1)\right]\nonumber\\
	=& D_i^{a+1}+\rho D_p^{a+1}+\rho\tilde{\Delta}_{a+2}(1).
\end{align*}

Substituting the above equation iteratively into the expression of $\tilde{\pi}$ in Eq.~\eqref{eq:thresboundsmall1} with $\tilde{\pi}=\tilde{\pi}_a^{*,\delta}+\lambda(1-\tilde{\pi}_a^{*,\delta})$, we prove the lower bound. 

\section*{Proof of Theorem \ref{theo:oracle}}
According to "stricter is better" assumption in Eq.~\eqref{assump:severerbetter}, the optimal policy when $\theta_{t}=1$ before $T$ is action $a_t=A$. Since the agent would avoid paying intervention cost when change-point has not occurred, the optimal policy when $\theta_{t}=1$ and when $t\geq T$ should be $a_t=0$. Therefore, our oracle policy in Theorem~\ref{theo:oracle} achieves the lowest cost.

We first consider the total propagation cost $C_p^{\mu_o}$. By the definition of the oracle policy, the distribution of $z_t$ is $\boldsymbol{\alpha}$ all the time, therefore, we have:
\begin{align*}
	C_p^{\mu_o}
	=&\sum_{T=1}^\infty \rho^{T-1}(1-\rho)\sum_{t=1}^{T-1}\left(\sum_{z=1}^Z\alpha(z)c_p^z\right)\nonumber\\
	=& \left(\sum_{z=1}^Z\alpha(z)c_p^z\right)\sum_{T=1}^\infty\rho^{T-1}(1-\rho)(T-1)\nonumber\\
	=& \frac{\rho}{1-\rho}\left(\sum_{z=1}^Z\alpha(z)c_p^z\right).
\end{align*}

We then consider the intervention cost $C_i^{\mu_o}$. By definition of oracle policy, the intervention cost only occurs when $\tau-1\leq t\leq T-2$ where the intervention level is not $0$. Therefore, we have:
\begin{align*}
	C_i^{\mu_o}=&\sum_{T=1}^\infty\rho^{T-1}(1-\rho)\sum_{\tau=1}^\infty\lambda(1-\lambda)^{\tau-1}\left(\sum_{t=0}^{T-2}c_i^A 1_{t\geq\tau-1}\right)\nonumber\\
	=&\sum_{T=1}^\infty\rho^{T-1}(1-\rho)\sum_{t=0}^{T-2}\left(\sum_{\tau=1}^{t+1}\lambda(1-\lambda)^{\tau-1}c_i^A\right)\nonumber\\
	=&\sum_{T=1}^\infty\rho^{T-1}(1-\rho)\sum_{t=0}^{T-2}c_i^A\left(1-(1-\lambda)^{t+1}\right)\nonumber\\
	=&c_i^A\sum_{T=1}^\infty\rho^{T-1}(1-\rho)\left[T-\frac{1}{\lambda}+\frac{(1-\lambda)^{T}}{\lambda}\right]\nonumber\\
	=&c_i^A\left(\frac{\rho}{1-\rho}-\frac{\rho(1-\lambda)}{1-\rho(1-\lambda)}\right).
\end{align*}

Combining $C_p^{\mu_o}$ and $C_i^{\mu_o}$, we get the result and prove the total cost of oracle policy.
	
\end{document}